\documentclass[11pt]{article}

\usepackage{epsfig}
\usepackage{latexsym}
\usepackage{graphicx}
\usepackage{color}
\usepackage{amsmath,amssymb}
\usepackage{cite}

\makeatletter
\def\section{\@startsection {section}{1}{\z@}{-3.5ex plus -1ex minus -.2ex}{2.3ex plus .2ex}{\large\bf}}
\def\subsection{\@startsection{subsection}{2}{\z@}{-3.25ex plus -1ex
minus -.2ex}{1.5ex plus .2ex}{\normalsize\bf}}
\makeatother
\newcommand{\captionfonts}{\small}
\makeatletter  
\long\def\@makecaption#1#2{%
  \vskip\abovecaptionskip
  \sbox\@tempboxa{{\captionfonts #1: #2}}%
  \ifdim \wd\@tempboxa >\hsize
    {\captionfonts #1: #2\par}
  \else
    \hbox to\hsize{\hfil\box\@tempboxa\hfil}%
  \fi
  \vskip\belowcaptionskip}
\makeatother   

\catcode`@=11
\def\marginnote#1{}
\newcount\hour
\newcount\minute
\newtoks\amorpm
\hour=\time\divide\hour
by60
\minute=\time{\multiply\hour by60 \global\advance\minute
by-\hour}
\edef\standardtime{{\ifnum\hour<12 \global\amorpm={am}
\else\global\amorpm={pm}\advance\hour by-12 \fi
 \ifnum\hour=0
\hour=12 \fi
 \number\hour:\ifnum\minute<10
0\fi\number\minute\the\amorpm}}
\edef\militarytime{\number\hour:\ifnum\minute<10
0\fi\number\minute}
\def\draftlabel#1{{\@bsphack\if@filesw
{\let\thepage\relax
 \xdef\@gtempa{\write\@auxout{\string
\newlabel{#1}{{\@currentlabel}{\thepage}}}}}\@gtempa
 \if@nobreak
\ifvmode\nobreak\fi\fi\fi\@esphack}
\gdef\@eqnlabel{#1}}
\def\@eqnlabel{}
\def\@vacuum{}
\def\draftmarginnote#1{\marginpar{\raggedright\scriptsize\tt#1}}
\def\draft{\oddsidemargin
0.0truein
 \def\@oddfoot{\sl preliminary draft \hfil
\rm\thepage\hfil\sl\today\quad\militarytime}
 \let\@evenfoot\@oddfoot
\overfullrule 3pt
 \let\label=\draftlabel
\let\marginnote=\draftmarginnote
\def\@eqnnum{(\theequation)\rlap{\kern\marginparsep\tt\@eqnlabel}
\global\let\@eqnlabel\@vacuum}
}
\catcode`@=12

\newcommand{\beq}{\begin{eqnarray}}
\newcommand{\eeq}{\end{eqnarray}}

\def\mchi01{m_{\tilde{\chi}^0_1}}
\def\mst1{m_{\tilde{t}_1}}
\def\msc1{m_{\tilde{c}_L}}
\def\mn1{m_{\tilde{\chi}_1^0}}

\def\acdj1{a_{j1}^{\tilde{c} d_k}}
\def\atdj1{a_{j1}^{\tilde{t} d_k}}
\def\bcdj1{b_{j1}^{\tilde{c} d_k}}
\def\btdj1{b_{j1}^{\tilde{t} d_k}}
\def\mcN{\mathcal N}

\newcommand{\s}{\newline \vspace*{-3.5mm}}
\newcommand{\id}{{\rm 1\kern-.12em
\rule{0.3pt}{1.5ex}\raisebox{0.0ex}{\rule{0.1em}{0.3pt}}}}
\renewcommand{\thefootnote}{\fnsymbol{footnote}}
\begin{document}

\thispagestyle{empty}

\begin{center}
\hfill CERN-PH-TH/2012-180 \\
\hfill KA-TP-25-2012 \\
\hfill SFB/CPP-12-42

\begin{center}

\vspace{1.7cm}

{\LARGE\bf Higgs Boson Masses in the Complex NMSSM \\[0.1cm] at
  One-Loop Level} 
\end{center}

\vspace{1.4cm}

{\bf T. Graf}\footnote{Present email address:
  thorben\_graf@gmx.de}, {\bf R. Gr\"ober$^{\,a}$}, {\bf
  M. M\"uhlleitner$^{\,a}$}, {\bf H. Rzehak}\footnote{On leave from:
  Albert-Ludwigs-Universit\"at Freiburg, Physikalisches Institut,
  Freiburg, Germany.}{\bf $^{\,b}$} and {\bf K. Walz$^{\,a}$} \\

\vspace{1.2cm}

${}^a\!\!$
{\em {Institut f\"ur Theoretische Physik, Karlsruhe Institute of Technology, 76128 Karlsruhe, Germany}
}\\
${}^b\!\!$
{\em {CERN, PH-TH, 1211 Geneva 23, Switzerland}}

\end{center}

\vspace{0.8cm}

\centerline{\bf Abstract}
\vspace{2 mm}
\begin{quote}
\small
The Next-to-Minimal Supersymmetric Extension of the Standard Model
(NMSSM) with a Higgs sector containing five neutral and two charged
Higgs bosons allows for a rich phenomenology. In addition, the plethora
of parameters provides many sources of CP violation. In contrast to the
Minimal Supersymmetric Extension, CP violation in the Higgs sector is already possible at tree-level. For a reliable understanding and interpretation of the experimental results of the Higgs boson search, and for a proper distinction of Higgs sectors provided by the Standard Model or possible extensions, the Higgs boson masses have to be known as precisely as possible including higher-order corrections. In this paper we calculate the one-loop corrections to the neutral Higgs boson masses in the complex NMSSM in a Feynman diagrammatic approach adopting a mixed renormalization scheme based on on-shell and $\overline{\mbox{DR}}$ conditions. We study various scenarios where we allow for tree-level CP-violating phases in the Higgs sector and where we also study radiatively induced CP violation due to a non-vanishing phase of the trilinear coupling $A_t$ in the stop sector. The effects on the Higgs boson phenomenology are found to be significant. We furthermore estimate the theoretical error due to unknown higher-order corrections by both varying the renormalization scheme of the top and bottom quark masses and by adopting different renormalization scales. The residual theoretical error can be estimated to about 10\%.
\end{quote}

\newpage
\renewcommand{\thefootnote}{\arabic{footnote}}
\setcounter{footnote}{0}

\section{Introduction}
The search for the Higgs boson and ultimately the understanding of the
mechanism behind the creation of particle masses represents one of
the major goals of the Large Hadron Collider (LHC). Recently, the
experimental collaborations ATLAS and CMS have updated their
results on the search for the Higgs boson. Both experiments observe
an excess of events  in the low Higgs mass range, compatible with a
Standard Model (SM) Higgs boson mass hypothesis close to 124 GeV at
3.1$\sigma$ local significance as reported by CMS \cite{cmshiggs} and
close to 126 GeV at 3.5$\sigma$ local significance in the ATLAS experiment
\cite{atlashiggs}.
 This is still too far away from the 5$\sigma$
required to claim discovery and necessitates the accumulation of
further data in the ongoing experiment. The slight excess of events in the
$\gamma\gamma$ final state signature as compared to the Standard Model expectation 
may hint to
the existence of new physics. One of the most popular extensions of
the SM are supersymmetric models (SUSY) \cite{susy}. While the Higgs sector
of the Minimal Supersymmetric Extension (MSSM) \cite{mssm} consists of
two complex Higgs doublets, which lead to five physical Higgs states
after electroweak symmetry breaking (EWSB), the Next-to-Minimal
Supersymmetric Model (NMSSM) \cite{nmssm} extends the Higgs sector by
an additional singlet superfield $\hat{S}$. This entails 7 Higgs bosons after EWSB,
which in the limit of the real NMSSM can be divided into three neutral purely
CP-even, two neutral purely CP-odd and two charged Higgs bosons, and in total
leads to five neutralinos. Although more complicated than the MSSM, the NMSSM
has several attractive features. Thus it allows for the dynamical
solution of the $\mu$ problem \cite{muproblem} through the scalar
component of the singlet field acquiring a non-vanishing vacuum
expectation value. Furthermore, the tree-level mass value of the
lightest Higgs boson is increased by new contributions to the quartic
coupling so that the radiative corrections necessary to shift the
Higgs mass to $\sim 125$ GeV are less important than in the MSSM
allowing for lighter stop masses\footnote{The bulk of the radiative
  corrections stems from the (s)top loops.} and less finetuning
\cite{finetune,King:2012is,brchanges0}.  The enlarged Higgs and
neutralino sectors, finally, 
lead to a richer phenomenology both in collider and dark matter (DM)
experiments. The latter is due to the possibility of a singlino-like
lightest neutralino, the former due to heavier Higgs
bosons decaying into lighter ones at sizeable rates or due to possibly
enhanced or suppressed branching ratios in LHC standard search
channels such as $\gamma\gamma$ or vector boson final states
\cite{brchanges0,brchanges}, to cite 
only a few of the possible modifications compared to SM or MSSM
phenomenology.  \s

The enlarged parameter set in supersymmetric theories allows for
further sources of CP violation as compared to the SM, where the only
source of CP violation occurs in the CKM matrix. Hence, the soft SUSY
breaking couplings and gaugino masses as well as the Higgsino mixing parameter
$\mu$ can be complex. While in the MSSM CP violation in the Higgs
sector is not possible at tree-level due to the minimality conditions
of the Higgs potential, it can be radiatively induced through
non-vanishing CP phases \cite{radcpmssm0,radcpmssm,1lfull}. Consequently, the
CP-even and CP-odd Higgs bosons mix so that the physical states have no
definite CP quantum number any more leading to substantial modifications in
Higgs boson phenomenology \cite{mssmcpveff}. The Higgs couplings to
the SM gauge bosons and fermions, their SUSY partners and the Higgs
self-couplings can be considerably modified compared to the CP-conserving case
inducing significant changes in the Higgs boson production rates and
decay modes. This could allow for Higgs bosons with masses below
present exclusion bounds from LEP and possibly Tevatron and LHC as
they might have escaped detection due to suppressed couplings involved
in the various standard Higgs search channels, which would then
necessitate new search strategies \cite{escapedet}. \s

In the NMSSM CP violation in the Higgs sector is possible at
tree-level. Though spontaneous tree-level CP violation in the
$Z_3$-invariant NMSSM is impossible due to vacuum stability
\cite{spontcpviol}, explicit CP violation can be realized already at
tree-level in contrast to the MSSM. In principle, there can be six
complex phases parametrizing the CP violation in the Higgs sector, two
relative phases between the vacuum expectation values (VEVs) of the
Higgs doublet and singlet fields and four
phases for the complex parameters
$\lambda,\kappa,A_\lambda,A_\kappa$. At tree-level these phases appear only in certain 
combinations, however, and exploiting tadpole conditions we are left with only one
independent phase combination. Explicit CP violation in the
Higgs sector leads to potentially large corrections to the electric
dipole moments (EDMs). The non-observation of EDMs for thallium, neutron and
mercury \cite{cpviolconstr} severely constrains the CP-violating
phases. However, as the phase combinations occurring in the EDMs can be
different from the ones inducing Higgs mixing, the phases can be
chosen such that the contributions to the EDMs are small while the
phases important for the Higgs sector can still be sizeable \cite{cpphaseok}.
This provides additional CP violation necessary for electroweak
baryogenesis \cite{ewbg}.
The explicit tree-level CP violation induces scalar-pseudoscalar
mixings between the doublet fields $H_{u,d}$ and the singlet field $S$, 
but not between the scalar and pseudoscalar components of the Higgs
doublets $H_{u,d}$. The latter is realized in scenarios where explicit
CP violation in the Higgs sector is induced through radiative
corrections. Radiative CP 
violation furthermore allows for a moderate amount of CP violation
which is still in agreement with the bounds from the EDMs
\cite{garisto}. In this respect, the CP
phases which play a role are $\varphi_{A_t}, \varphi_{A_b}$ from the
trilinear couplings $A_t,A_b$. They are involved in the dominant
corrections from the third generation squark loops. Phases from third
generation Yukawa couplings on the other hand can be reabsorbed by
redefinitions of the quark fields when neglecting generation mixing.
 Further sources for radiative CP
violation stem from the gaugino sector where the soft SUSY breaking mass parameters
$M_1,M_2$ and $M_3$ are in general complex. One of the two parameters
$M_1$ and $M_2$ can be chosen real by applying an $R$-symmetry
transformation.  
The gluino mass parameter $M_3$ and hence its phase enters only at
the two-loop level.
\s

Radiatively induced CP-violating effects from the third generation
squark sector have been considered in the effective potential approach
at one-loop level in Refs.~\cite{squarkrad}. One-loop contributions from the
charged particle loops have been taken into account by
Refs.~\cite{radall}, also in the effective potential approach. In
Ref.~\cite{Funakubo:2004ka} the third generation (s)quark and gauge
contributions are included in the one-loop effective potential. The
full one-loop and logarithmically enhanced two-loop effects in the
renormalization-group improved approach have been included in 
\cite{cpeff2loop}.
In order to properly interpret the results from the experiments and
distinguish the various Higgs sectors from each other, a precise
knowledge of the Higgs boson masses and couplings at the highest
possible accuracy, including higher-order corrections, is
indispensable. In this paper we consider the full one-loop corrections
to the Higgs boson masses in the CP-violating NMSSM in the Feynman
diagrammatic approach.\footnote{Higher-order corrections to the Higgs
  boson masses in the real NMSSM can be found in
  Refs.\cite{horealnmssm,realnmssm}.} We allow for explicit CP violation at
tree-level 
by including non-vanishing CP phases for the Higgs doublets and
singlet, as well as for $\lambda$, $\kappa$, $A_\lambda$ and $A_\kappa$, which
effectively reduce to one physical CP phase combination at tree-level when the 
tadpole conditions are exploited. We furthermore allow for radiatively induced CP
violation stemming from the stop\footnote{For the low values of
  $\tan\beta$ applied in our numerical analysis, the CP-violating
  effects from the sbottom sector are marginal.} by choosing a
non-vanishing CP phase for $A_t$. The Higgs
sector as well as the neutralino and chargino sector will be used to
determine the counterterms. The renormalization is performed
in a mixed scheme which
combines $\overline{\mbox{DR}}$ conditions for the
parameters not directly related to physical observables 
with on-shell (OS) conditions for the physical input
values. 
For the choice of our parameter sets the recent
constraints from the Higgs 
boson searches at LEP \cite{LEPHbb}, Tevatron \cite{tevsearch} and LHC
\cite{cmshiggs,atlashiggs} are taken
into account. The inclusion of CP violation affects the Higgs
phenomenology and hence the validity of possible scenarios. 
Our results will therefore help to
clarify the question what kind of Higgs sector may be realized
in nature for a SM-like Higgs boson with mass around 125 GeV, should
the tantalizing hints of the LHC experiments be confirmed by the
discovery of the Higgs boson at the 5$\sigma$ level once a sufficient
amount of data is accumulated. \s 

The organization of the paper is as follows. In Sect.~\ref{sec:Complex} the parameters of the complex NMSSM will be
introduced. Section \ref{sec:Higgs} presents the details of our
calculation. After introduction of the complex tree-level Higgs sector
in Sect.~\ref{sec:tree} the set of input parameters is given in
Sect.~\ref{sec:input}. 
The renormalization conditions and determination of
the Higgs masses as well as the mixing angles are discussed in
Sects.~\ref{sec:one-loop} and \ref{sec:oneloopmass}, respectively. In 
Sect.~\ref{sec:numerical} we present our numerical analysis, discuss the
influence of different renormalization schemes as well as the
consequences of one-loop corrections in the complex NMSSM for the Higgs
boson phenomenology and results at the LHC. We terminate with the
conclusions in Sect.~\ref{sec:summary}.

\section{\label{sec:Complex} Complex parameters in the NMSSM}

The Lagrangian of the complex NMSSM can be divided into an MSSM part which is
adopted from the MSSM Lagrangian  and an additional NMSSM part. The
latter contains (apart from the phases of 
the Higgs doublets and singlets) 
four
 additional complex parameters.  
Two of them are 
the coupling $\kappa$ of the self-interaction
of the new singlet superfield $\hat{S}$ and the coupling $\lambda$ for
the interaction of $\hat{S}$ with the Higgs doublet superfields
$\hat{H}_{u}$ and $\hat{H}_d$ ($\hat{H}_{u}$ and $\hat{H}_d$  couple to the
up- and down-type quark superfields, respectively). They are introduced
via the extension of the MSSM superpotential $W_{\text{MSSM}}$, 
\begin{align} 
W_{\text{NMSSM}} = W_{\text{MSSM}} - \epsilon_{ab} \lambda \hat{S} \hat{H}_d^a
\hat{H}_u^b + \frac{1}{3} \kappa \hat{S}^3~,
\end{align}
with $\epsilon_{12} = \epsilon^{12} =1$. 
The MSSM part of the superpotential is given by
\begin{align}
W_{\text{MSSM}} = - \epsilon_{ab} \bigl( y_u \hat{H}_u^a \hat{Q}^b \hat{U}^c -
y_d \hat{H}_d^a \hat{Q}^b \hat{D}^c - y_e \hat{H}_d^a \hat{L}^b
\hat{E}^c\bigr)
\; ,
\end{align}
where $\hat{Q}$ and $\hat{L}$ are the left-handed quark and lepton
superfield doublets and $\hat{U}$, $\hat{D}$ and $\hat{E}$ are the 
right-handed  
up-type, down-type and electron-type superfield singlets, respectively.  
The superscript $c$ denotes charge 
conjugation. Colour and generation indices have been  
omitted. The quark and lepton Yukawa couplings are given by $y_d$,
$y_u$ and $y_e$.  They are in general complex. However, when neglecting
generation mixing, as we assume in this paper, the phases of the
Yukawa couplings can be reabsorbed by redefining the quark fields \textit{i.e.} 
these phases can be chosen arbitrarily without changing the physical 
meaning~\cite{CKM}.

The soft SUSY breaking Lagrangian in the NMSSM is also extended with
respect to the MSSM,
\begin{align}
\mathcal L_{\text{NMSSM}}^{\text{soft}} = \mathcal
L_{\text{MSSM}}^{\text{soft}} - m_S^2 |S|^2 
   + 
 (\epsilon_{ab} A_{\lambda} \lambda S H_d^a H_u^b - \frac{1}{3}
   A_{\kappa}\kappa S^3 + h.c.%
) \; ,
\end{align}
 containing two further complex parameters specific to
the NMSSM, the soft 
SUSY breaking trilinear
couplings $A_\lambda$ and $A_\kappa$.
The MSSM part is given by\footnote{Here the indices of the soft SUSY
  breaking masses, $Q$ ($L$) stand for the
  left-handed doublet of the three quark (lepton) generations and $U,D,E$
  are the indices for the right-handed up-type and down-type fermions and charged
  leptons, respectively. In the trilinear coupling parameters
  the indices $u,d,e$ represent the up-type and down-type fermions and charged
  leptons.}
\begin{align}\nonumber
\mathcal L_{\text{MSSM}}^{\text{soft}} &= - m_{H_d}^2 |H_d|^2 - m_{H_u}^2 
|H_u|^2  
- m_{Q}^2 |\tilde{Q}|^2  - m_{U}^2 |\tilde{u}_R|^2  - m_{D}^2 |\tilde{d}_R|^2 
- m_{L}^2 |\tilde{L}|^2 - m_{E}^2 |\tilde{e}_R|^2 \\& \quad \nonumber
+ \epsilon_{ab} ( y_u A_u H_u^a \tilde{Q}^b \tilde{u}^*_R 
                -  y_d A_d H_d^a \tilde{Q}^b \tilde{d}^*_R 
                -  y_e A_e H_d^a \tilde{Q}^b \tilde{e}^*_R + h.c.)\\& \quad
- \frac{1}{2} (M_1  \tilde{B} \tilde{B} +   M_2 \tilde{W}_i \tilde{W}_i + M_3
\tilde{G} \tilde{G} + h.c) \; .
\end{align}
The soft SUSY breaking trilinear couplings $A_u$, $A_d$, $A_e$ of
the up-type, 
down-type and charged lepton-type sfermions\footnote{We
  neglect generation mixings so that we have nine complex numbers
  $A_u$, $A_d$, $A_e$ instead of three complex $3\times 3$~matrices.}, 
respectively, which are
already present in the MSSM, are in general complex. However, the soft SUSY breaking mass parameters
of the scalar fields, $m_{X}^2$ ($X=S$, $H_d$, $H_u$, $Q$, $U$, $D$, $L$, $E$),
are real. The SM-type and 
SUSY fields forming a superfield (denoted with a hat) are represented 
by a letter without and with a tilde,
respectively: $\tilde{Q}$, $\tilde{L}$ and $\tilde{u}_R$, $\tilde{d}_R$,
$\tilde{e}_R$ are the  superpartner fields corresponding to the
left- and right-handed quark and lepton fields. 
In general, also the soft SUSY breaking mass parameters of the gauginos,
$M_1$, $M_2$ and $M_3$, are complex 
where the gaugino fields are denoted by $\tilde{B}$, $\tilde{W}_i$
($i=1,2,3$) and $\tilde{G}$ for the bino, the winos and the gluinos
corresponding to the weak hypercharge $U(1)$, the weak isospin $SU(2)$
and the colour $SU(3)$ symmetry. The $R$-symmetry can then be
exploited to choose either $M_1$ or $M_2$ to be real. 
The kinetic and gauge interaction part of the NMSSM
Lagrangian finally do not contain any complex parameters.\s

Expressing the Higgs boson fields as an expansion about the vacuum expectation
values, two further phases appear, 
\begin{align}
H_d = \begin{pmatrix}\frac{1}{\sqrt{2}}(v_d + h_d + i a_d) \\ 
                      h_d^- \end{pmatrix}~, \;\;\; 
H_u = e^{i \varphi_u} \begin{pmatrix} h_u^+ \\
            \frac{1}{\sqrt{2}}(v_u + h_u + i a_u) \end{pmatrix}~, \;\;\; 
S = \frac{e^{i \varphi_s}}{\sqrt{2}} (v_s + h_s + i a_s)~.
\label{Higgsdecomp}
\end{align}
The phases $\varphi_u$ and $\varphi_s$ describe the phase differences
between the three vacuum expectation values $\langle H_d^0\rangle$, $\langle H_u^0\rangle$ and $\langle S\rangle$. In case of vanishing phases, $\varphi_u = \varphi_s = 0$, the fields $h_i$ and  $a_i$ with 
$i = d, u, s$  correspond to the CP-even and CP-odd part of the
neutral entries of $H_u$, $H_d$ and $S$. The charged components are denoted
by $h_i^\pm$ ($i = d, u$). 

Exploiting that the phases of the Yukawa couplings can be chosen arbitrarily,
the phase of the up-type coupling is set to 
 $\varphi_{y_u} = -
  \varphi_u$ while the down-type and the charged lepton-type ones are assumed to be 
real. This choice ensures that the quark and lepton mass terms yield real
masses without any further phase transformation of the corresponding fields. 
 \s

In the following  renormalization procedure we will make use
of the chargino and neutralino sectors, therefore they are  introduced briefly here.
The fermionic superpartners of the neutral Higgs bosons and colourless gauge
bosons are $\tilde{H}^0_d$ and $\tilde{H}^0_u$ for 
the neutral components of the Higgs doublets, $\tilde{S}$ for the Higgs
singlet, the bino $\tilde{B}$ and
 the neutral component $\tilde{W}_3$ of the winos. After electroweak
 symmetry breaking these fields mix, and in the Weyl spinor basis
 $\psi^0 =  (\tilde{B},\tilde{W}_3, \tilde{H}^0_d,\tilde{H}^0_u,  \tilde{S})^T$ 
the neutralino mass matrix $M_N$ can be written as 
\begin{align}
&M_N = \nonumber \\& \begin{pmatrix} 
M_1               & 0        & - c_\beta M_Z s_{\theta_W} &   
                                 M_Z s_\beta s_{\theta_W} e^{-i \varphi_u}
               & 0\\
0                 & M_2      &    c_\beta M_W    & - M_W s_\beta e^{-i \varphi_u} 
               & 0\\
- c_\beta M_Z s_{\theta_W} & c_\beta M_W & 0                &
                 - \lambda \frac{v_s}{\sqrt{2}} e^{i \varphi_s}   & 
                - \frac{\sqrt{2} M_W s_\beta s_{\theta_W} \lambda e^{i\varphi_u}}{e}
  \\
 M_Z s_\beta s_{\theta_W} e^{-i \varphi_u} &  - M_W s_\beta e^{-i \varphi_u} & 
                            - \lambda \frac{v_s}{\sqrt{2}}e^{i \varphi_s} & 0 &
                          -  \frac{\sqrt{2} M_W c_\beta s_{\theta_W} \lambda}{e} 
   \\
0   & 0     & -  \frac{\sqrt{2}M_W s_\beta s_{\theta_W}\lambda e^{i \varphi_u}}{e}  & 
                            - \frac{\sqrt{2} M_W c_\beta s_{\theta_W}\lambda}{e} &
                             \sqrt{2} \kappa v_s e^{i \varphi_s}
\end{pmatrix}        
\end{align}
where $M_W$ and $M_Z$ are the $W$ and $Z$ boson masses, respectively. 
The angle
$\beta$ is defined via the ratio of the vacuum expectation values of the two
Higgs doublets, $\tan \beta = v_u/v_d$, $\theta_W$ denotes the
electroweak mixing angle and $e$ is the electric charge. 
From here on the short hand
notation $c_x = \cos x$, $s_x = \sin x$ and $t_x = \tan x$ is used. 

The neutralino mass matrix $M_N$ is complex\footnote{Note, that in general the parameters
$\lambda$, $\kappa$, $M_1$ and $M_2$  are complex.} but symmetric and can be
diagonalized with the help of the $5 \times 5$ matrix $\mathcal N$, yielding  
$\text{diag}(m_{\tilde{\chi}^0_1},
m_{\tilde{\chi}^0_2},m_{\tilde{\chi}^0_3},m_{\tilde{\chi}^0_4},
m_{\tilde{\chi}^0_5})
= \mathcal N^* M_N \mathcal N^\dagger$,  
where the absolute mass values are ordered as $|m_{\tilde{\chi}^0_1}|\leq ... 
\leq
|m_{\tilde{\chi}^0_5}|$. The neutralino mass eigenstates $\tilde{\chi}^0_i$, expressed 
as a Majorana spinor, can
then be obtained by 
\begin{align}\label{eq:neuspinor}
\tilde{\chi}^0_i = (\chi^0_i, \overline{\chi^0_i})^T \quad\ \text{with}\quad\ \chi^0_i = \mathcal N_{ij} \psi^0_j,\quad\  i,\,j = 1,\dots,5~.
\end{align}

The fermionic superpartners of the charged Higgs and gauge bosons are
given in terms of the Weyl spinors $\tilde{H}_d^\pm$,
$\tilde{H}_u^\pm$, $\tilde{W}_1$ and $\tilde{W}_2$ where the latter
two can be reexpressed as  $\tilde{W}^\pm = (\tilde{W}_1 \mp i
\tilde{W}_2)/\sqrt{2}$. Arranging these Weyl spinors as
\begin{align}\label{eq:chaspinor}
\psi^-_R = \begin{pmatrix} \tilde{W}^- \\ \tilde{H}_d^- \end{pmatrix}, \quad\
\psi^+_L = \begin{pmatrix} \tilde{W}^+ \\ \tilde{H}_u^+ \end{pmatrix}
\end{align}
leads to mass terms of the form, $(\psi^-_R)^T M_C \psi^+_L + h.c. $, with the 
chargino mass matrix  
\begin{align}
M_C = \begin{pmatrix} M_2 & \sqrt{2}  s_\beta M_W e^{-i \varphi_u}\\
   \sqrt{2} c_\beta M_W & \lambda \frac{v_s}{\sqrt{2}} e^{i \varphi_s}
    \end{pmatrix}~.
\end{align}
The chargino mass matrix can be diagonalized with the help of two unitary 
$2 \times 2$ matrices, $U$ and $V$, yielding
$\text{diag}(m_{\tilde{\chi}^\pm_1}, m_{\tilde{\chi}^\pm_2}) = U^* M_C
V^\dagger$ with $m_{\tilde{\chi}^\pm_1} \leq  m_{\tilde{\chi}^\pm_2}$. The
left-handed and the right-handed part of the mass eigenstates are 
\begin{align}
\tilde{\chi}^+_L = V \psi^+_L~, \quad \tilde{\chi}^-_R = U \psi^-_R~,
\end{align}
respectively, with the mass eigenstates $\tilde{\chi}^+_i =
(\tilde{\chi}^+_{L_i}, \overline{\tilde{\chi}^-_{R_i}})^T$, $i=1,2$, written as a Dirac spinor.

\section{\label{sec:Higgs} The Higgs Boson Sector in the Complex NMSSM}

\subsection{\label{sec:tree} The Higgs Boson Sector at  Tree-Level}

To ensure the minimum of 
the Higgs potential $V_{\text{Higgs}}$ at
non-vanishing vacuum expectation values $v_u$, $v_d$, $v_s$ the terms
linear in the Higgs boson fields have to vanish according to
\begin{align}\label{taddef}
t_{\phi} \equiv \frac{\partial V_{\text{Higgs}}}{\partial \phi}|_{\text{Min.}} \stackrel{!}{=} 0 \quad\ \text{for}\quad\
 \phi = h_d, h_u, h_s, a_d, a_u, a_s.
\end{align}
At tree-level, these tadpole parameters $t_{\phi}$ are given by\footnote{The complex
  parameters are expressed in terms of their absolute value and a
  complex phase, {\it i.e.} for example $\lambda \equiv |\lambda|
  e^{i\varphi_\lambda}$.}
 \begin{align}
t_{h_d} &=\bigl[m_{H_d}^2 + 
   \frac{M_Z^2 c_{2 \beta}}{2}
- v_s t_\beta |\lambda| (\frac{|A_{\lambda}|}{\sqrt{2}} c_{\varphi_x} +
|\kappa| \frac{v_s}{2} c_{\varphi_y}) 
+ |\lambda|^2  (\frac{2  s_\beta^2 M_W^2 s_{\theta_W}^2}{e^2} +
\frac{v_s^2}{2})\bigr] \frac{2 c_\beta M_W s_{\theta_W}}{e}~,\label{tadhd}
\\
t_{h_u} &=\bigl[m_{H_u}^2 - \frac{M_Z^2
    c_{2\beta}}{2}  
 - \frac{|\lambda|  v_s}{t_\beta}  (  \frac{|A_{\lambda}|}{\sqrt{2}}
 c_{\varphi_x}+|\kappa| \frac{v_s}{2}  c_{\varphi_y})
+  |\lambda|^2 (\frac{2 c_\beta^2 M_W^2 s_{\theta_W}^2}{e^2} +
\frac{v_s^2}{2})\bigr]\frac{2 s_\beta M_W s_{\theta_W}}{e}~,
\\
t_{h_s} &= m_S^2 v_s  - 
\Bigl[ s_{2 \beta} |\lambda|(\frac{|A_{\lambda}|}{\sqrt{2}}
c_{\varphi_x} +|\kappa| v_s c_{\varphi_y}) 
- |\lambda|^2  v_s\Bigr] \frac{2  M_W^2 s_{\theta_W}^2}{e^2} +  |\kappa|^2 v_s^3  +   
 \frac{1}{\sqrt{2}} |A_{\kappa}| |\kappa| v_s^2 c_{\varphi_z}~, \label{tadhs}
\\
t_{a_d} &=    \frac{M_W s_{\theta_W} s_\beta}{e} |\lambda| v_s
   (\sqrt{2} |A_{\lambda}| s_{\varphi_x} - |\kappa| v_s 
s_{\varphi_y}) ~,\label{tadad}\\
t_{a_u} &= 
 \frac{1}{t_\beta} t_{a_d}~,\\
t_{a_s} &= \frac{2 M_W^2 s_{\theta_W}^2 s_{2 \beta}}{e^2} |\lambda|(
\frac{1}{\sqrt{2}}|A_{\lambda}| s_{\varphi_x} + |\kappa| v_s s_{\varphi_y} )
 -\frac{1}{\sqrt{2}} |A_{\kappa}| |\kappa| v_s^2 s_{\varphi_z} \;  , \label{tadas}
\end{align}
where we have introduced a short hand notation for the following phase combinations
\begin{align}
\varphi_x &= \varphi_{A_\lambda} +
\varphi_{\lambda} + \varphi_s + \varphi_u~, \label{eq:varphix}\\[0.1cm]
 \varphi_y &= \varphi_{\kappa} - \varphi_{\lambda} + 2 \varphi_s -
 \varphi_u~, \label{eq:varphiy} \\[0.1cm]
\varphi_z &= \varphi_{A_{\kappa}} + \varphi_{\kappa} + 3 \varphi_s~.\label{eq:varphiz}
\end{align}
In the expressions for the tadpole parameters some of the original parameters have
already been replaced in favour of the parameters on which we will impose 
our renormalization conditions, as
described in detail in Sects.~\ref{sec:input} and
\ref{sec:one-loop}.
Thus the vacuum expectation
values $v_u$, $v_d$ and the $U(1)$ and $SU(2)$ gauge couplings $g'$
and $g$ have been replaced by 
$\tan \beta = v_u/v_d$, the gauge boson
masses $M_W$ and $M_Z$ and the electric charge $e$ (according to
Eqs.~\eqref{Ersetzung2} and \eqref{Ersetzung3} in Appendix~\ref{sec:trafo}). This replacement has also been applied in the expressions of the mass matrices given below.

It should be noted that the Eqs.~\eqref{tadad} and \eqref{tadas} can have zero, one or 
two solutions for $\varphi_x,\varphi_z \in [-\pi, \pi)$ depending on the values of the parameters. 
If no solution is found there is no minimum of the Higgs potential at the corresponding set of
values $v_d, v_u, v_s$ and thus this parameter point is discarded. The single solutions yield one of the
two values $\varphi_x,\varphi_z = \pm \pi/2$. 
Assuming there exist two solutions of Eq.~\eqref{tadad} 
then if $\varphi_x^S$ with $\varphi_x^S>0$ solves this equation 
then also $\pi -\varphi_x^S$ is a solution and similarly if $\varphi_x^S$ with $\varphi_x^S<0$ is a solution then 
$-(\pi -\varphi_x^S)$ is the second solution, analogously for $\varphi_z^S$.

The terms of the Higgs potential which are bilinear in the neutral
Higgs boson fields contribute to the corresponding $6 \times 6$ Higgs 
boson mass matrix $M_{\phi\phi}$ in the basis of $\phi = (h_d, h_u, h_s, a_d, a_u, a_s)^T$
which can be expressed in terms of three $3 \times 3$ matrices
$M_{hh}$,  $M_{aa}$ and $M_{ha}$ 
\begin{align}
M_{\phi\phi} = \begin{pmatrix} M_{hh}&  M_{ha}\\
                          M_{ha}^T &
                          M_{aa}\end{pmatrix} 
\label{eq:higgsmassmatrix}
\end{align}
where $M_{hh}$ and $M_{aa}$ are symmetric matrices. 

The
entries of $M_{hh}$ describing the mixing of the CP-even 
 components of the Higgs doublet and singlet fields read
\begin{align}
M_{h_dh_d} &=  M_Z^2 c_\beta^2 
+ \frac{1}{2} |\lambda| v_s t_\beta (\sqrt{2}|A_{\lambda}| c_{\varphi_x} + |\kappa| v_s c_{\varphi_y})~,\\
M_{h_dh_u} &= -\frac{1}{2} M_Z^2 s_{2 \beta} 
- \frac{1}{2} |\lambda| v_s (\sqrt{2}|A_{\lambda}| c_{\varphi_x} + |\kappa| v_s c_{\varphi_y})
+ 2 |\lambda|^2 \frac{M_W^2 s_{\theta_W}^2}{e^2} s_{2 \beta}~,\\
M_{h_uh_u} &=  M_Z^2 s_\beta^2
+ \frac{1}{2} |\lambda| \frac{v_s}{t_\beta}(\sqrt{2}|A_{\lambda}| c_{\varphi_x} + |\kappa| v_s c_{\varphi_y})~,\\
M_{h_dh_s} &= 2 |\lambda|^2 \frac{M_W s_{\theta_W}}{e} c_\beta v_s 
- |\lambda|  \frac{M_W s_{\theta_W}}{e} s_\beta  (\sqrt{2}|A_{\lambda}| c_{\varphi_x} + 2|\kappa| v_s c_{\varphi_y})~,\\
M_{h_uh_s} &= 2 |\lambda|^2 \frac{M_W s_{\theta_W}}{e} s_\beta v_s 
-  |\lambda|\frac{M_W s_{\theta_W}}{e} c_\beta (\sqrt{2}|A_{\lambda}| c_{\varphi_x} + 2|\kappa| v_s c_{\varphi_y})~,\\
M_{h_sh_s} &= 2 |\kappa|^2 v_s^2 + \frac{v_s}{\sqrt{2}} |\kappa| |A_{\kappa}| c_{\varphi_z} 
+ \sqrt{2}|A_\lambda| |\lambda|\frac{M_W^2 s^2_{\theta_W}}{e^2 v_s} s_{2 \beta} c_{\varphi_x}~.
\end{align}
Note, that here the tadpole conditions
Eqs.~\eqref{tadhd}--\eqref{tadhs} together with Eq.~\eqref{taddef} have already been applied to 
eliminate $m_{H_d}^2$, $m_{H_u}^2$ and $m_S^2$. Exploiting additionally
Eqs.~\eqref{tadad} and \eqref{tadas} we can eliminate $c_{\varphi_x}$ and
$c_{\varphi_z}$ through 
\begin{align}
c_{\varphi_x} &= \pm \sqrt{1 - \frac{|\kappa|^2 v_s^2}{2|A_\lambda|^2}
  s^2_{\varphi_y}}~, \label{eq:cphixsign} \\
c_{\varphi_z}  &= \pm \sqrt{1 - 18 \frac{M_W^4 s_{\theta_W}^4
    s_{2\beta}^2 |\lambda|^2}{e^4 |A_{\kappa}|^2 v_s^2}
  s^2_{\varphi_y}}~. \label{eq:cphizsign} 
\end{align}
The two signs correspond to the two possible solutions of Eqs.~\eqref{tadad} and \eqref{tadas} (as explained before). 
Choosing either solution will define the sign of
the cosine. In our numerical evaluation we will treat the sign as a further input. 

The mixing of the CP-odd components of the Higgs doublet and singlet
fields is characterized by the matrix $M_{aa}$ which has the 
following entries,
\begin{align}
M_{a_da_d} &= \frac{1}{2}|\lambda|(\sqrt{2}|A_{\lambda}| c_{\varphi_x} +
|\kappa| v_s c_{\varphi_y})v_s t_\beta \; ,\\
M_{a_da_u} &=  \frac{1}{2}|\lambda| (\sqrt{2}|A_{\lambda}| c_{\varphi_x}
+ |\kappa| v_s c_{\varphi_y}) v_s  \; ,\\
M_{a_ua_u} &= \frac{1}{2}|\lambda|(\sqrt{2}|A_{\lambda}| c_{\varphi_x} + |\kappa| v_s c_{\varphi_y})\frac{v_s}{t_\beta}
 \; , \\
M_{a_da_s} &= |\lambda| \frac{M_W s_{\theta_W}}{e} s_\beta (\sqrt{2}|A_{\lambda}| c_{\varphi_x} 
- 2 |\kappa| v_s c_{\varphi_y})  \; ,
\end{align}
\begin{align}
M_{a_ua_s} &= |\lambda| \frac{M_W s_{\theta_W}}{e}
c_\beta(\sqrt{2}|A_{\lambda}| c_{\varphi_x} - 2 |\kappa| v_s
c_{\varphi_y})  \; ,\\
M_{a_sa_s} &= |\lambda| (\sqrt{2}|A_{\lambda}| c_{\varphi_x}+ 4 |\kappa| v_s c_{\varphi_y})
\frac{M_W^2 s_{\theta_W}^2}{e^2 v_s} s_{2\beta}
 - 3 |A_\kappa| |\kappa| \frac{v_s}{\sqrt{2}} c_{\varphi_z} \; , 
\end{align}
where again Eq.~\eqref{taddef} together with the Eqs.~\eqref{tadhd}--\eqref{tadhs} have been applied and
Eqs.~\eqref{taddef}, \eqref{tadad} and \eqref{tadas} can be used to  replace $c_{\varphi_x}$ and 
$c_{\varphi_z}$. The matrix $M_{ha}$ governs the mixing between
the CP-even and the CP-odd  components of the Higgs doublet and
singlet fields,
\begin{align}
M_{ha} = \begin{pmatrix} 0      & 0             & 3 v_s s_\beta \\
                                0      & 0             & 3 v_s c_\beta \\
                 - v_s s_\beta & - v_s c_\beta & - 4 s_{2 \beta} \frac{M_W s_{\theta_W}}{e}
              \end{pmatrix} \frac{M_W s_{\theta_W}}{e}
              |\kappa||\lambda|s_{\varphi_y} \;.
\end{align}
In case of $\varphi_y = n_y \pi$, $n_y \in {\mathbb Z}$, the entries of $M_{ha}$ vanish and hence, in that case, there is no CP violation
at tree-level in the NMSSM Higgs sector;  
after transformation to the mass eigenstates we are left with
three purely CP-even and two purely CP-odd Higgs bosons. \s

The transformation into mass eigenstates can be performed in two steps. 
In our approach, first, the Goldstone boson field is extracted by applying the 
$6 \times 6$ rotation matrix\footnote{The explicit form of $\mathcal
  R^{G}$ can be found in Appendix~\ref{sec:Higgsdetails}.} $\mathcal R^{G}$,
\begin{align}
\Phi_i = \mathcal R^{G}_{ij} \phi_j \; , 
\end{align}
where 
$\Phi = (h_d, h_u, h_s, A, a_s, G)^T$. 
The resulting mass matrix, 
\begin{align}
M_{\Phi\Phi} = \mathcal R^G M_{\phi\phi} {\mathcal R^G}^T~,
\end{align}
 can finally be diagonalized with the help of the matrix 
$\mathcal R$ leading to 
\begin{align}
 \mathcal R M_{\Phi\Phi} {\mathcal R}^T = 
\text{diag}\Bigl((M_{H_1}^{(0)})^2, ..., (M_{H_5}^{(0)})^2, 0\Bigr) =: \mathcal{D}_H \label{eq:rndef}
\end{align}
 with the mass values being
ordered as $M_{H_1}^{(0)} \leq ... \leq M_{H_5}^{(0)}$ and the superscript $(0)$ denoting the tree-level values of the masses. 
The corresponding mass
eigenstates are obtained as 
\begin{align}
H_i = \mathcal R_{ij} \Phi_j~.
\end{align}

The mass matrix $M_{h^+\,h^-}$ of the charged entries of the Higgs
doublet fields, 
\begin{align}
(h^+_d, h^+_u) M_{h^+\,h^-}(h^-_d, h^-_u)^T~,\
\end{align}
 is explicitly given as
\begin{align} M_{h^+\,h^-} =  \frac{1}{2} \begin{pmatrix} t_\beta & 1 \\ 1 & \frac{1}{t_\beta} \end{pmatrix} 
\Bigl[M_W^2 s_{2 \beta} + |\lambda|v_s(\sqrt{2}|A_{\lambda}| c_{\varphi_x} + |\kappa| v_s c_{\varphi_y}) 
 - 2|\lambda|^2 \frac{M_W^2 s^2_{\theta_W}}{e^2} s_{2\beta}\Bigr] \; ,
\end{align}
where again the Eqs.~\eqref{taddef}--\eqref{tadhs} have already been applied.
Diagonalizing this mass matrix with the help of a rotation matrix with the angle $\beta_c$,
where $\beta_c = \beta$ at tree-level, 
yields the mass of the physical charged Higgs boson,
\begin{align}
M_{H^\pm}^2 =   M_W^2 + \frac{|\lambda|v_s}{s_{2 \beta}} (\sqrt{2}|A_{\lambda}| c_{\varphi_x} + |\kappa| v_s c_{\varphi_y}) 
 - 2|\lambda|^2 \frac{M_W^2 s^2_{\theta_W}}{e^2} \;,
\end{align}
 and a mass of zero for the charged Goldstone boson.

\subsection{\label{sec:input} Set of Input Parameters for the Higgs Boson Sector}
To summarize, the original parameters entering the Higgs potential and thereby also the Higgs mass matrix are 
\begin{align}
m_{H_d}^2, m_{H_u}^2, m_S^2, \varphi_{A_{\kappa}}, \varphi_{A_{\lambda}},
|A_\lambda|, g, g', v_u, v_d,  v_s, \varphi_s, \varphi_u, 
 |\lambda|, \varphi_{\lambda}, |\kappa|,
\varphi_{\kappa}, |A_{\kappa}|~.\label{eq:orgparset}
\end{align}
Instead of using this set of original parameters it is convenient to convert it to a set of parameters which offer an intuitive interpretation. This is especially true for the parameters which can be replaced by gauge boson masses squared as they are measurable quantities. We have chosen the set,
\begin{align}
\underbrace{t_{h_d}, t_{h_u}, t_{h_s}, t_{a_d}, t_{a_s}, M_{H^\pm}^2, M_W^2, M_Z^2, e}_{\text{on-shell}}, 
\underbrace{ \tan \beta,   v_s, \varphi_s, \varphi_u, |\lambda|,
  \varphi_{\lambda}, |\kappa|,  
\varphi_{\kappa}, |A_{\kappa}|}_{\overline{\text{DR}}}~ 
\label{eq:defparset}
\end{align}
where the first part of the parameters are directly related to ``physical'' quantities\footnote{Whether the tadpole parameters can be called physical quantities is debatable but certainly their introduction is motivated by physical interpretation. Therefore, in a slight abuse of the language, we are also calling the renormalization conditions for the tadpole parameters on-shell.} and will be defined via on-shell conditions while the remaining parameters are understood as $\overline{\text{DR}}$~parameters (see Sect.~\ref{sec:parren}). The transformation rules for going from set 
Eq.~\eqref{eq:orgparset} to 
set Eq.~\eqref{eq:defparset} are given in Appendix~\ref{sec:trafo}.

\subsection{\label{sec:one-loop} The Higgs Boson Sector at One-Loop Level}
At one-loop level, the Higgs boson sector and the corresponding relations between parameters of the theory and physical quantities are changed by radiative corrections. In particular, the Higgs boson mass matrix receives contributions from the renormalized self-energies\footnote{In general, we call $\hat\Sigma$ and $\Sigma$ renormalized and unrenormalized self-energy, respectively.}
 $\hat\Sigma_{ij}(p^2)$ at an external momentum squared $p^2$,
\beq 
\hat{\Sigma}_{ij}(p^2)=\Sigma_{ij}(p^2)+\frac{1}{2}p^2\left[\delta \mathcal{Z}^{\dagger}+\delta \mathcal{Z}\right]_{ij}-\frac{1}{2}\left[\delta \mathcal{Z}^{\dagger}\mathcal{D}_{H}+\mathcal{D}_{H}^{\dagger}\delta \mathcal{Z}\right]_{ij}-[
\mathcal{R} \delta \mathcal{M}_{\Phi\Phi}\mathcal{R}^{\dagger}]_{ij}~, \label{eq:HiggsSE}
\eeq
with $i, j = 1, \dots, 6$ and $H_6 = G$ the Goldstone boson. 
The 
unrenormalized self-energies $\Sigma_{ij}$ are obtained by taking into account all possible contributions to the Higgs boson self-energy, 
including the ones from fermion, gauge boson, Goldstone boson, Higgs boson, chargino, neutralino, sfermion and ghost loops. 

The wave function renormalization matrix $\delta\mathcal{Z}$ in the basis of the Higgs boson mass eigenstates is derived  via rotation from the 
corresponding matrix $\delta \mathcal{Z}_{\Phi}$ in the basis of the Higgs boson states $\Phi$,
\begin{align}
\delta\mathcal{Z}=\mathcal{R} \delta \mathcal{Z}_{\Phi}\mathcal{R}^\dagger~.
\end{align}
The Higgs boson fields $\Phi$ are renormalized by replacing the fields by renormalized ones and a renormalization factor. This can be expressed as, valid up to one-loop order, with the field renormalization constant
 $\delta \mathcal{Z}_{\Phi}$ as 
\begin{align}
\Phi \rightarrow (\id + \frac{1}{2} \delta \mathcal{Z}_{\Phi}) \Phi~,
\end{align}
 where $\delta\mathcal{Z}_{\Phi} = \mathcal R^G \delta \mathcal{Z}_{\phi} \mathcal R^{G^\dagger}$. The field renormalization constant $\delta \mathcal{Z}^{\phi}$ of the interaction eigenstates $\phi$
is a diagonal matrix
\begin{align}\label{eq:dZIA}
\delta\mathcal{Z}_{\phi}= \text{diag}(
\delta Z_{H_u},\,\delta Z_{H_d},\,\delta Z_{S},\delta Z_{H_d},\,\delta Z_{H_u},\,\delta Z_{S})~.  
\end{align}                                                                                                                               
The explicit definitions and expressions for $\delta Z_{H_u}$,  $\delta Z_{H_d}$ and  $\delta Z_{S}$ are given in Sect.~\ref{sec:Zdef}.

The matrix $\delta \mathcal{M}_{\Phi \Phi}$ denotes the counterterm matrix in the basis of the Higgs boson states~$\Phi$ which has been introduced within the 
renormalization procedure by replacing the parameters in the mass matrix as given in Appendix \ref{sec:trafo} in Eqs.~\eqref{Ersetzung1}--\eqref{Ersetzung3} by their renormalized ones and corresponding counterterms and expanding about
 the renormalized parameters. The part linear in the counterterms forms the mass matrix counterterm. The specific definitions of the parameters and 
the determination of the counterterms are discussed in Sect.~\ref{sec:parren}.

\subsubsection{\label{sec:Zdef} Higgs boson field renormalization}
The field renormalization constants introduced in Eq.~\eqref{eq:dZIA} are defined in the
$\overline{\text{DR}}$ scheme. The precise expressions for $\delta
Z_{H_d}$, $\delta Z_{H_u}$ and $\delta Z_{S}$ are determined via the
following system of equations
\begin{align}\nonumber
\delta Z_{H_d} &(|\mathcal R_{i1}|^2  + |\mathcal R_{i4} \sin\beta +
\mathcal R_{i6} \cos\beta|^2) +
\delta Z_{H_u} (|\mathcal R_{i2}|^2  + |\mathcal R_{i4} \cos\beta - \mathcal
R_{i6} \sin\beta|^2) \\ &+
 \delta Z_{S}  (|\mathcal R_{i3}|^2  + |\mathcal R_{i5}|^2) =
- \Sigma_{ii}'|_{\text{div}} \quad
\text{with} \quad i = 1,2,3~,  \label{Zfaktorglg}
\end{align}
where 
\beq\Sigma_{ii}'=\frac{\partial \Sigma_{ii}(p^2)}{\partial p^2}\Big|_{p^2
= (M_{H_i}^{(0)})^2}~.
\eeq
The subscript ${\text{'div'}}$ 
denotes that only the divergent parts proportional to $\Delta$ are taken into account with $\Delta = 2/(4 - D) - \gamma_E + \ln 4 \pi$  and  $\gamma_E$ being the Euler constant and $D$
 the number of dimensions. The pole of $\Delta$  for $D = 4$ characterizes the divergences. 
Solving this system of equations Eq.~\eqref{Zfaktorglg} yields
\begin{align}
\delta Z_{H_d} &=\frac{1}{r}\bigl[(r_{23} r_{32} - r_{22} r_{33})  \Sigma_{11}'
 + (r_{12} r_{33} - r_{13} r_{32} )  \Sigma_{22}'
+ (r_{13} r_{22} -  r_{12} r_{23})   \Sigma_{33}'\bigr]_{\text{div}} \label{eq:dZHd}~,\\
\delta Z_{H_u} &=\frac{1}{r}\bigl[ (r_{21} r_{33} - r_{23} r_{31}) \Sigma_{11}'
 + (r_{13} r_{31} - r_{11} r_{33}) \Sigma_{22}' 
+ (r_{11} r_{23} -  r_{13} r_{21})  \Sigma_{33}'\bigr]_{\text{div}} \label{eq:dZHu}~,\\
\delta Z_{S} &=\frac{1}{r}\bigl[(r_{22} r_{31} - r_{21} r_{32}) \Sigma_{11}'
 + (r_{11} r_{32} - r_{12} r_{31}) \Sigma_{22}' 
+ (r_{12} r_{21} -  r_{11} r_{22})  \Sigma_{33}'\bigr]_{\text{div}}
\label{eq:dZS}~,
\end{align}
with
\begin{align}
r_{i1} &= (|\mathcal R_{i1}|^2  + |\mathcal R_{i4} \sin\beta + \mathcal R_{i6} \cos\beta|^2) \label{eq:ri1}~,\\
r_{i2} &= (|\mathcal R_{i2}|^2  + |\mathcal R_{i4} \cos\beta - \mathcal R_{i6} \sin\beta|^2) \label{eq:ri2}~,\\
r_{i3} &= (|\mathcal R_{i3}|^2  + |\mathcal R_{i5}|^2)~,\\
r &= r_{11} r_{22} r_{33}  + r_{12} r_{23} r_{31} + r_{13} r_{32} r_{21}  - r_{11} r_{23} r_{32} - r_{13} r_{22} r_{31} -  r_{12} r_{21} r_{33} ~.
\end{align}
It should be noted that $\mathcal R_{i6} = 0$ for $i \neq 6$ and hence, in Eqs.~\eqref{eq:ri1} and \eqref{eq:ri2} terms proportional to 
$\mathcal R_{i6}$ vanish for the values $i = 1,\,2,\,3$ needed in Eqs.~\eqref{eq:dZHd}--\eqref{eq:dZS}.

\subsubsection{\label{sec:parren} Parameter renormalization}
The parameter renormalization is performed by replacing the parameters by the renormalized ones and the corresponding counterterms,
\begin{align}
t_{\phi} &\rightarrow t_{\phi} + \delta t_{\phi} \quad \hspace*{1.5cm}\text{with} \quad \phi = \{h_d, h_u, h_s, a_d, a_s\}~,\\
M_{H^\pm}^2 &\rightarrow M_{H^\pm}^2 + \delta M_{H^\pm}^2~, \quad\ 
 M_W^2 \rightarrow  M_W^2  + \delta M_W^2~, \quad\ 
 M_Z^2 \rightarrow  M_Z^2  + \delta M_Z^2~, \\
e &\rightarrow (1 + \delta Z_e) e~,\\
\tan \beta &\rightarrow \tan \beta + \delta \tan \beta~, \quad v_s  \rightarrow v_s + \delta v_s~,\\
 \varphi_s &\rightarrow \varphi_s + \delta \varphi_s~, \quad \hspace*{1cm}\varphi_u \rightarrow \varphi_u + \delta \varphi_u~,\\
\lambda &\rightarrow \lambda + \delta \lambda = 
\lambda + e^{i\varphi_\lambda} \delta |\lambda| + i \lambda \,\delta \varphi_\lambda~, \quad
\kappa \rightarrow \kappa + \delta \kappa = \kappa + e^{i\varphi_\kappa} \delta |\kappa| + i \kappa \,\delta \varphi_\kappa~,
\label{eq:dladkap}\\
|A_{\kappa}|&\rightarrow |A_{\kappa}| +\delta |A_{\kappa}|~.
\end{align}
In the case of complex parameters the complex counterterms can be understood in terms of two real counterterms, one for the absolute value and 
one for the phase, as in Eq.~\eqref{eq:dladkap} for $\delta \lambda$ and $\delta \kappa$.

As we make use of the chargino and the neutralino sector for the determination of the counterterms $\delta v_s$, $\delta \varphi_s$, $\delta \lambda$, 
$\delta\kappa$ and $\delta \varphi_u$ we also need to renormalize the gaugino mass parameters $M_1$ and $M_2$,
\begin{align}
M_1 &\rightarrow M_1 + \delta M_1 = M_1 + e^{i \varphi_{M_1}}\delta |M_1| + i M_1 \delta \varphi_{M_1}~, \\ 
M_2 &\rightarrow M_2 + \delta M_2 = M_2 + e^{i \varphi_{M_2}}\delta |M_2| + i M_2 \delta \varphi_{M_2}~.
\end{align}
To keep the relations as general as possible we do not make use of the $R$-symmetry relations here and keep both gaugino mass parameters complex.

In the following, we list all the renormalization conditions and counterterms. The renorma\-li\-za\-tion scheme applied here is a generalization of the
``mixed scheme'' of Ref.~\cite{realnmssm} for complex parameters -- we will be brief on the conditions that can be directly taken from
 Ref.~\cite{realnmssm}.

\begin{itemize}
\item[(i-v)] \underline{\it Tadpole parameters:}\\[1ex]
The renormalization conditions for the tadpole parameters are chosen such that the linear terms of the Higgs boson 
fields in the Higgs potential also vanish at one-loop level,
\begin{align}
\delta t_\phi = T_\phi \quad \text{with} \quad \phi = h_d, h_u, h_s, a_d, a_s~,
\end{align}
where $T_\phi$ denotes the contribution of the irreducible one-loop tadpole diagrams.
\item[(vi - viii)]  \underline{\it Masses of the gauge bosons and the charged
  Higgs boson:}\\[1ex]
The masses of the gauge bosons and of the charged Higgs boson are determined via on-shell conditions requiring that the
mass parameters squared correspond to the pole masses squared leading to
\begin{align}
\delta M_W^2 = \widetilde{\text{Re}}\Sigma^T_{WW} (M_W^2)~, \quad \delta M_Z^2 = \widetilde{\text{Re}}\Sigma^T_{ZZ} (M_Z^2)~, \quad 
\delta M_{H^\pm}^2 = \widetilde{\text{Re}}\Sigma_{H^\mp H^\pm} (M_{H^\pm}^2)~,
\end{align}
where $\Sigma^T_{WW}$ and $\Sigma^T_{ZZ}$ are the transverse parts of the unrenormalized $W$ boson and $Z$ boson self-energy, 
respectively, while $\Sigma_{H^\mp H^\pm}$ denotes the unrenormalized self-energy of the charged Higgs boson. $\widetilde{\text{Re}}$ takes only the
real part of the scalar loop functions but keeps the complex structure of the parameters.
\item[(ix)]  \underline{\it Electric charge:}\\[1ex]
The electric charge is fixed via the $e\bar{e}\gamma$ vertex in such a way that this vertex does not receive any corrections 
at the one-loop level in the Thomson limit, \textit{i.e.} for zero
momentum transfer. This yields ({\it cf.}~Ref.~\cite{Denner:1991kt} up to a
different sign convention in the second term)
\begin{align}
\delta Z_e = \frac{1}{2}{\Sigma^T}'_{\!\!\!\!\gamma\gamma}(0) + \frac{s_{\theta_W}}{
  c_{\theta_W} M_Z^2} \Sigma^T_{\gamma Z}(0)~, 
\end{align} 
with $\Sigma^T_{\gamma\gamma}$ and $\Sigma^T_{\gamma Z}$ being the transverse part of the unrenormalized  photon self-energy and the  
unrenormalized mixing of photon and $Z$~boson, respectively.
%
\item[(x)] \underline{\it Ratio of the vacuum expectation values $\tan
  \beta$:}\\[1ex] 
The ratio of the vacuum expectation values $\tan \beta$ is defined as a
$\overline{\text{DR}}$ parameter and the counterterm is given by \cite{vuvd,Frank:2003tg}
\begin{align}
\delta \tan \beta = \frac{\tan \beta}{2}\bigl[\delta Z_{H_u} - \delta Z_{H_d}\bigr]|_{\text{\scriptsize div}}~.
\end{align}
\item[(xi,xii)] \underline{\it Vacuum expectation value $v_s$ and phase
  $\varphi_s$:}\\[1ex]
The singlet vacuum expectation value $v_s$ and the phase $\varphi_s$ are determined as 
$\overline{\text{DR}}$ parameters. 
For the derivation of the corresponding counterterms, we start out from the on-shell conditions for the chargino masses,
\begin{align}
\widetilde{\text{Re}}\hat{\Sigma}_{\chi^+_{ii}} (p)\; \tilde{\chi}^+_i (p) |_{p^2 = m_{\chi^\pm_i}^2} = 0, \quad i = 1,\,2,
\end{align} 
where $\hat{\Sigma}_{\chi^+_{ii}}$ are the renormalized chargino self-energies. Applying the 
 decomposition of the fermionic self-energy 
\begin{align}
\hat{\Sigma}_{ij} (p^2) = \slash{\!\!\! p} \hat{\Sigma}^L_{ij} (p^2) {\cal P}_L + 
\slash{\!\!\! p} \hat{\Sigma}^R_{ij} (p^2) {\cal P}_R + 
 \hat{\Sigma}^{Ls}_{ij} (p^2) {\cal P}_L + \hat{\Sigma}^{Rs}_{ij} (p^2) {\cal P}_R 
\label{eq:strucself}
\end{align}
with ${\cal P}_{L,R} = (\id \mp \gamma_5)/2$ being the left- and
right-handed projectors, leads to the finite relations 
\begin{align}
\bigl[m_{\tilde{\chi}^\pm_i} \bigl(\widetilde{\text{Re}}\hat{\Sigma}^L_{\chi^\pm} (p^2) +
                                 \widetilde{\text{Re}}\hat{\Sigma}^R_{\chi^\pm} (p^2) \bigr)
                            +  \widetilde{\text{Re}}\hat{\Sigma}^{Ls}_{\chi^\pm} (p^2) +
                                \widetilde{\text{Re}}\hat{\Sigma}^{Rs}_{\chi^\pm} (p^2)
  ]_{ii}  &= 0 \;,
\label{eq:vsrel1}\\
\bigl[m_{\tilde{\chi}^\pm_i} \bigl( \widetilde{\text{Re}}\hat{\Sigma}^L_{\chi^\pm} (p^2) -
                                  \widetilde{\text{Re}}\hat{\Sigma}^R_{\chi^\pm} (p^2) \bigr)
                            - \widetilde{\text{Re}}\hat{\Sigma}^{Ls}_{\chi^\pm} (p^2) +
                                   \widetilde{\text{Re}}\hat{\Sigma}^{Rs}_{\chi^\pm} (p^2)
         \bigr]_{ii} &= 0 \:,
\label{eq:vsrel2}
\end{align}
which can be exploited for the determination of the counterterms.
 Using the expressions
given in Eqs.~\eqref{eq:cren1}--\eqref{eq:cren4} in Appendix~\ref{sec:renselfChaNeu}  for
the renormalized self-energies yields
\begin{align}
\text{Re}(U^* \delta M_C V^\dagger)|_{\text{\scriptsize div}} &= \frac{1}{2}
                  \bigl[m_{\tilde{\chi}^\pm_i}\bigl({\Sigma}^L_{\chi^+} (p^2) + 
                                {\Sigma}^R_{\chi^+} (p^2)\bigr) +
                        {\Sigma}^{Ls}_{\chi^+} (p^2) + {\Sigma}^{Rs}_{\chi^+}
                        (p^2) 
                       \bigr]_{ii}|_{\text{\scriptsize div}} \nonumber\\&=:
                  \text{Re}\,\delta m_{\chi^+_{ii}}~,\label{eq:chare}\\
\text{Im}(U^* \delta M_C V^\dagger)|_{\text{\scriptsize div}} &= \frac{i}{2}\bigl[
 {\Sigma}^{Rs}_{\chi^+} (p^2) - {\Sigma}^{Ls}_{\chi^+}(p^2) 
+i m_{\tilde{\chi}^\pm_i} (U^*\, \text{Im} \delta Z^C_R U^T +
 V\, \text{Im} \delta Z^C_L V^\dagger)  \bigr]_{ii} |_{\text{\scriptsize div}} \nonumber\\&=:
                  \text{Im}\,\delta m_{\chi^+_{ii}}\label{eq:chaim} \:.
\end{align}
The imaginary parts of the field renormalization constants have been set to zero, 
hence $\text{Im} \delta Z^C_R = \text{Im} \delta Z^C_L = 0$.
With
\begin{align}
\delta M_C = \begin{pmatrix} \delta M_2 & 
\sqrt{2} e^{-i \varphi_u} [  \delta (M_W s_\beta) 
- i  s_\beta M_W \delta \varphi_u]\\[0.1cm]
   \sqrt{2}\, \delta (c_\beta M_W) & 
 \frac{e^{i \varphi_s}}{\sqrt{2}} [\lambda \delta v_s
   + i \lambda v_s \delta \varphi_s + 
 v_s \delta \lambda]
    \end{pmatrix}\label{eq:dMC} \; ,
\end{align}
solving Eqs.~\eqref{eq:chare} and \eqref{eq:chaim} for  $\delta v_s + i v_s \delta \varphi_s$ and 
$\delta M_2$\footnote{Even though $M_2$ does
not enter the Higgs boson sector at tree-level, it has to be dealt with due to its entanglement in Eq.~\eqref{eq:dMC}.}, we obtain
\begin{align}
\delta M_2 &= 
\frac{1}{|U_{11}|^2 - |V_{12}|^2}\bigr[
  V_{11} U_{11} \delta m_{\chi^+_{11}} 
- V_{21} U_{21} \delta m_{\chi^+_{22}} \nonumber \\[0.2cm]& \qquad\qquad\qquad\qquad
- U_{11} U_{12}^* \delta M_{C_{21}}|_{\text{\scriptsize div}} 
- V_{11} V_{12}^* \delta M_{C_{12}}|_{\text{\scriptsize div}}\bigr]
\label{eq:M2}~,\\
\delta v_s + i v_s \delta \varphi_s 
 &= \frac{\sqrt{2} \lambda^* e^{-i
    \varphi_s}}{|\lambda|^2(|U_{11}|^2 - |V_{12}|^2)}\bigr[
-  V_{12} U_{12} \, \delta m_{\chi^+_{11}} 
+ V_{22} U_{22}  \, \delta m_{\chi^+_{22}} \nonumber \\[0.2cm]&  \qquad\qquad\qquad\qquad
+ U_{11}^* U_{12}  \,  
\delta M_{C_{12}}|_{\text{\scriptsize div}} 
+ V_{11}^* V_{12}  \, 
\delta M_{C_{21}}|_{\text{\scriptsize div}}\bigr]-
 v_s \lambda^* \frac{\delta \lambda}{|\lambda|^2}~.
\label{eq:vs}\end{align} 
It should be noted that $\delta \varphi_u$ contained in 
$\delta M_{C_{12}}$ as well as $\delta \lambda$ 
have not been defined yet. We need additional conditions given by Eqs.~\eqref{eq:lambda}--\eqref{eq:phiu}. Together
with Eqs.~\eqref{eq:M2} and \eqref{eq:vs} they form a
system of linear equations that can be easily solved but leads to lengthy expressions.
\item[(xiii -- xvii)] \underline{\it Couplings $\lambda$ and $\kappa$ and the
  phase $\varphi_u$:}\\[1ex]
The  phase $\varphi_u$ as well as $\lambda$ and $\kappa$
  are also defined as $\overline{\text{DR}}$ parameters. 
On-shell conditions for the neutralino masses\footnote{In terms of an on-shell scheme, if both chargino masses were defined on-shell only three of the neutralino masses could be chosen independently. Although Eq.~\eqref{eq:neuos} leads to two independent equations for each $i$, for $i = 4$ it is only partly used to fix one still undefined phase.} 
\begin{align}\label{eq:neuos}
\widetilde{\text{Re}}\hat{\Sigma}_{\chi^0_{ii}} (p) \, \tilde{\chi}^0_i (p) |_{p^2 = m_{\chi^0_i}^2} = 0, \quad i = 1,\dots, 4~,
\end{align} 
 are exploited to derive the 
following finite relations,\footnote{It should be noted that, in general, 
both Eqs.~\eqref{eq:neurel1} and
\eqref{eq:neurel2} for $i = 4$ only hold for the divergent part
simultaneously as all the parameters are already fixed by other
 conditions.}
\begin{align}
\bigl[m_{\tilde{\chi}^0_i} \bigl(\widetilde{\text{Re}}\hat{\Sigma}^L_{\chi^0} (p^2) +
                                 \widetilde{\text{Re}}\hat{\Sigma}^R_{\chi^0} (p^2) \bigr)
                            +  \widetilde{\text{Re}}\hat{\Sigma}^{Ls}_{\chi^0} (p^2) +
                               \widetilde{\text{Re}} \hat{\Sigma}^{Rs}_{\chi^0} (p^2)
  ]_{ii}  &= 0 \;,
\label{eq:neurel1}\\
\bigl[m_{\tilde{\chi}^0_i} \bigl( \widetilde{\text{Re}}\hat{\Sigma}^L_{\chi^0} (p^2) -
                                  \widetilde{\text{Re}}\hat{\Sigma}^R_{\chi^0} (p^2) \bigr)
                            - \widetilde{\text{Re}}\hat{\Sigma}^{Ls}_{\chi^0} (p^2) +
                                  \widetilde{\text{Re}} \hat{\Sigma}^{Rs}_{\chi^0} (p^2)
         \bigr]_{ii} &= 0 ~,
\label{eq:neurel2}
\end{align}
with $i = 1,\dots,4$.
Applying the Eqs.~\eqref{eq:nren1}--\eqref{eq:nren4} in Appendix~\ref{sec:renselfChaNeu} leads to
\begin{align}
\text{Re}(\mathcal N^* \delta M_N \mathcal N^\dagger)_{ii}|_{\text{\scriptsize
    div}} &= \bigl[m_{\tilde{\chi}^0_i} {\Sigma}^L_{\chi^0} (p^2)  +
                        \frac{1}{2}\bigl({\Sigma}^{Ls}_{\chi^0} (p^2)   +
                        {\Sigma}^{Rs}_{\chi^0} (p^2) \bigr)
                       \bigr]_{ii}|_{\text{\scriptsize div}} =:
                  \text{Re}\,\delta m_{\chi^0_{ii}}~, \label{eq:neure}\\
\text{Im}(\mathcal N^* \delta M_N \mathcal N^\dagger)_{ii}|_{\text{\scriptsize
    div}} &= 
\frac{i}{2}\bigl[
 {\Sigma}^{Rs}_{\chi^0} (p^2) - {\Sigma}^{Ls}_{\chi^0}(p^2) 
+ 2 i m_{\tilde{\chi}^0_j} ( 
 \mathcal N^* \text{Im} \delta Z^{N} \mathcal N^\dagger)  \bigr]_{ii}
|_{\text{\scriptsize div}}
\nonumber \\
&=: 
                  \text{Im}\,\delta m_{\chi^0_{ii}}~, \label{eq:neuim}
\end{align}
$i = 1,\dots,4$ where already
 ${\Sigma}^L_{\chi^0}|_{ii}={\Sigma}^R_{\chi^0}|_{ii}$ has been used which is true due to
the Majorana character of the neutralinos. 
The imaginary part of $\delta Z^N$ has been set to zero,
 $\text{Im} \delta Z^N = 0$.
The elements of the mass matrix counterterm $\delta M_N$ are
derived as 
{\allowdisplaybreaks
\begin{align}
\delta M_{N_{11}} &= \delta M_1~, \label{eq:dMN11}\\
\delta M_{N_{22}} &= \delta M_2~,\\
\delta M_{N_{55}} &= \sqrt{2} e^{i \varphi_s}\bigl[v_s  \delta \kappa + \kappa 
     \delta v_s + i \kappa v_s  \delta \varphi_s\bigr]~,\\
\delta M_{N_{13}} &= 
 - M_Z s_{\theta_W} c_\beta^2 s_\beta \; \delta \tan \beta
+ \frac{c_\beta}{2 s_{\theta_W} M_Z} \bigl[\delta M_W^2 
- \delta M_Z^2\bigr]~,  \\
\delta M_{N_{14}} &= e^{-i \varphi_u} [\delta(s_\beta M_Z s_{\theta_W}) - i
  s_\beta M_Z s_{\theta_W} \delta \varphi_u]~, \\
\delta M_{N_{23}} &= \delta(c_\beta M_W)~,\\
\delta M_{N_{24}} &= - e^{-i \varphi_u} [\delta(s_\beta M_W) - i 
  s_\beta M_W \delta \varphi_u]~, \\
\delta M_{N_{34}} &= - \frac{1}{\sqrt{2}}e^{i \varphi_s} \bigl[v_s  \delta \lambda
  + \lambda \delta v_s + i \lambda v_s \delta \varphi_s\bigr]~,\\
\delta M_{N_{35}} &= - \sqrt{2} e^{i \varphi_u}\bigl[
 \lambda  \frac{\delta(s_\beta M_W s_{\theta_W})}{e} 
- \frac{s_\beta M_W s_{\theta_W}}{e}(\lambda \delta Z_e - \delta \lambda 
- i   \lambda  \delta
\varphi_u) \bigr]~, \\
\delta M_{N_{45}} &= - \sqrt{2}\bigl[
\lambda  \frac{\delta(c_\beta M_W s_{\theta_W})}{e} 
+ \frac{c_\beta M_W s_{\theta_W}}{e} (\delta \lambda - \lambda \delta Z_e)\bigr]~, \\
\delta M_{N_{33}} &=\delta M_{N_{44}} =\delta M_{N_{12}} =\delta M_{N_{15}}
= \delta M_{N_{25}} = 0 ~.\label{eq:dMNzero}
\end{align}
}
As the neutralino mass matrix is symmetric, this also holds for the counterterm mass matrix and therefore $\delta M_{N_{ij}} =\delta M_{N_{ji}}$. 
Rewriting Eqs.~\eqref{eq:neure} and~\eqref{eq:neuim} explicitly and solving for $\delta M_1$ results in the following set of equations,
\begin{align}
&\delta M_1 = \frac{\mathcal N_{11}^2}{|\mathcal N_{11}|^4}\Bigl[\delta
m_{\chi^0_{11}}
- 2 \mcN_{11}^*[\mcN_{13}^* \delta M_{N_{13}}+\mcN_{14}^* \delta M_{N_{14}}]
- 2 \mcN_{12}^*[\mcN_{13}^* \delta M_{N_{23}}+\mcN_{14}^* \delta M_{N_{24}}]
\nonumber\\& \qquad\qquad
- 2 \mcN_{13}^*[\mcN_{14}^* \delta M_{N_{34}}+\mcN_{15}^* \delta M_{N_{35}}]
- 2 \mcN_{14}^* \mcN_{15}^* \delta M_{N_{45}} \nonumber\\& \qquad\qquad
- (\mcN_{12}^*)^2 \delta M_2 - (\mcN_{15}^*)^2 \delta M_{N_{55}}~,
\label{eq:M1} \\
&2 \bigl[a_{214} \delta M_{N_{14}} +  a_{224} \delta M_{N_{24}}+ 2 a_{234} \delta M_{N_{34}}
+  a_{235} \delta M_{N_{35}}+  a_{245} \delta M_{N_{45}}\bigr] 
\nonumber \\& \quad
+  a_{222} \delta M_2 +  a_{255} \delta M_{N_{55}}
\nonumber\\& \quad
= (\mcN^*_{21})^2 \delta m_{\chi^0_{11}} - (\mcN^*_{11})^2 \delta m_{\chi^0_{22}} - 2 a_{213} \delta M_{N_{13}} -  2 a_{223} \delta M_{N_{23}}~,
 \label{eq:lambda}\\
&2 \bigl[a_{314} \delta M_{N_{14}} +  a_{324} \delta M_{N_{24}}+ 2 a_{334} \delta M_{N_{34}}
+  a_{335} \delta M_{N_{35}}+  a_{345} \delta M_{N_{45}}\bigr]\nonumber\\& \quad
+  a_{322} \delta M_2 +  a_{355} \delta M_{N_{55}}
\nonumber\\& \quad
= (\mcN^*_{31})^2 \delta m_{\chi^0_{11}} - (\mcN^*_{11})^2 \delta m_{\chi^0_{33}} - 2 a_{313} \delta M_{N_{13}} -  2 a_{323} \delta M_{N_{23}}~,
 \label{eq:kappa}\\
&\text{Im}\Big\{2 \bigl[a_{414} \delta M_{N_{14}} +  a_{424} \delta M_{N_{24}}+ 2 a_{434} \delta M_{N_{34}}
+  a_{435} \delta M_{N_{35}}+  a_{445} \delta M_{N_{45}}\bigr]\nonumber\\& \quad
+  a_{422} \delta M_2 +  a_{455} \delta M_{N_{55}}\Big\}
\nonumber\\& \quad
= \text{Im}\Big\{(\mcN^*_{41})^2 \delta m_{\chi^0_{11}} - (\mcN^*_{11})^2 \delta m_{\chi^0_{44}} - 2 a_{413} \delta M_{N_{13}} -  2 a_{423} \delta M_{N_{23}}\Big\}~,
 \label{eq:phiu}
\end{align}
where we have introduced the shorthand notation
\begin{align}
a_{ijk} &= (\mcN^*_{i1})^2 \mcN^*_{1j} \mcN^*_{1k} -  (\mcN^*_{11})^2 \mcN^*_{ij}\mcN^*_{ik}~. 
\end{align} 
As stated above, taking Eqs.~\eqref{eq:lambda}--\eqref{eq:phiu} together with Eqs.~\eqref{eq:M2} and \eqref{eq:vs} leads to a system of equations with 4 complex and
one real equation linear in the counterterms that has to be solved for $\delta M_2$, $\delta v_s$, $\delta \varphi_s$, $\delta \lambda$,
$\delta \kappa$ and $\delta \varphi_u$. 
\item[(xviii)] \underline{\it Absolute value of the singlet trilinear
  coupling $|A_\kappa|$:}\\[1ex]
The absolute value of the singlet trilinear coupling $|A_\kappa|$ is
determined as a $\overline{\text{DR}}$~parameter. The corresponding
counterterm is calculated using
\begin{align}
\mathcal R_{i5} \mathcal R_{j 5}\hat{\Sigma}_{i j}(M_{a_s a_s}) = 0 \; ,
\end{align}
which is equivalent to
\begin{align}
\delta M_{a_s a_s} = \mathcal R_{i5} \mathcal R_{j 5}\Sigma_{i j}(M_{a_s
  a_s}) \; ,
\end{align}
with $\delta M_{a_s a_s}$ depending on $\delta |A_\kappa|$.
Dropping the finite parts and solving for $\delta |A_\kappa|$ yields
\begin{align}\nonumber
\delta |A_\kappa| &= - \frac{\sqrt{2}}{|\kappa| v_s \bigl[3 c_{\varphi_z} - 3\,
t_{\varphi_z}\frac{f_2}{|A_\kappa|}\bigr]}
\Bigl\{
\mathcal R_{i5} \mathcal R_{j 5}\Sigma_{i j}(M_{a_s,a_s}) 
- \delta f_1   \\& \quad\
+ \frac{3}{\sqrt{2}} |A_\kappa| \bigl[ v_s c_{\varphi_z} \delta |\kappa| + |\kappa|  
c_{\varphi_z} \delta v_s\bigr] 
+ \frac{3}{\sqrt{2}} |\kappa| v_s t_{\varphi_z} 
\delta f_2
\Bigr\}_{\text{div}}
\end{align}
with
\begin{align}
f_1 &=  \bigl[M_{H^\pm}^2 - M_W^2 c_{\Delta \beta}^2\bigr] \frac{M_W^2 s_{\theta_W}^2 s^2_{2 \beta}}{e^2 v_s^2 c_{\Delta \beta}^2}
- \frac{M_W s_{\theta_W} s_{2 \beta} c_\beta c_{\beta_B}^2}{e v_s^2 c_{\Delta \beta}^2} \bigl[t_{h_u} + t_{\beta} t_{\beta_B}^2 t_{h_d} \bigr]
+ \frac{t_{h_s}}{v_s}
\nonumber \\&
\quad\ + |\lambda| M_W^2 \frac{s^2_{\theta_W} s_{2 \beta}}{e^2 v_s^2} 
      \bigl[2 |\lambda| M_W^2 \frac{s_{\theta_W}^2}{e^2} s_{2 \beta} + 3 |\kappa| v_s^2 c_{\varphi_y}\bigr]~,\\
f_2 &= \frac{\sqrt{2}}{v_s} \Bigl( \frac{2 M_W s_{\theta_W} c_\beta}{e |\kappa| v_s^2} t_{a_d} +
  \frac{3 |\lambda| M_W^2 s_{\theta_W}^2 s_{2\beta}}{e^2} s_{\varphi_y} -
\frac{1}{|\kappa| v_s} t_{a_s}\Bigr)~. 
\end{align}
The counterterms $\delta f_1$ and $\delta f_2$, which are functions of counterterms of the parameters defined as input in Eq.~\eqref{eq:defparset}, are determined by replacing the parameters by  
renormalized ones plus corresponding counterterms and expanding
 about the parameters. It has to be taken into account that in the expressions
$\Delta \beta = \beta - \beta_B$ only $\beta$ is treated within the renormalization procedure. The angle $\beta_B = \beta_c = \beta_n$ is the mixing angle of the 
charged Higgs bosons
and the angle extracting the Goldstone boson defined in Appendix \ref{sec:trafo}.
\end{itemize}

Following the approach above, it has been found that the counterterms $\delta \varphi_s$, $\delta \varphi_\lambda$, $\delta 
\varphi_\kappa$, $\delta \varphi_u$, $\delta \varphi_{M_1}$ and $\delta \varphi_{M_2}$ vanish. 
In that respect, it is interesting to note that Eqs.~\eqref{eq:chaim} and \eqref{eq:neuim} allow for a 
certain freedom of choice; some potentially divergent parts can be moved into 
$\text{Im} \delta Z^C_{L},\; \text{Im} \delta Z^C_{R} $ and $\text{Im} \delta Z^N$ which do not appear in the calculation of 
the Higgs boson masses as the charginos and neutralinos only enter through internal lines in the 
Feynman diagrams.

The derivation of the counterterms for $v_s$, $\varphi_s$, $\lambda$, $\kappa$ and
$\varphi_u$ presented above is not unique. In a second approach on-shell conditions for all the neutralino 
masses plus an additional condition from the chargino sector, Eq.~\eqref{eq:chaim} for $i =1$, have been
 exploited to calculate the $\overline{\text{DR}}$ counterterms leading to the same result and providing
 a good cross-check. A further possibility is to determine the counterterms within the Higgs boson sector 
only. We have also done that but this does not test the calculation at the same level as using 
conditions from the chargino and neutralino sector.
\subsection{\label{sec:oneloopmass} Loop Corrected Higgs Boson Masses and Mixing Matrix Elements}
The one-loop corrected scalar Higgs boson masses
squared are extracted numerically as the zeroes of the determinant of the
two-point vertex functions $\hat{\Gamma}$,
\beq
\hat\Gamma(p^2)=i\big(\id\cdot p^2-\mathcal{M}^{\text{1l}}\big)\qquad \text{with} \qquad 
\big(\mathcal{M}^{\text{1l}}\big)_{ij}= \big(M^{(0)}_{H_i}\big)^2\delta_{ij}-\hat\Sigma_{ij}(p^2)\quad i,\,j = 1,\dots,5~,
\label{eq:sematrix}
\eeq
where $\hat\Sigma_{ij}(p^2)$ is given in Eq.~\eqref{eq:HiggsSE}. The superscript $1l$ denotes the one-loop order. 
\par
Starting from Eq.~\eqref{eq:sematrix} the Higgs masses at one-loop level can be obtained via an iterative procedure.\footnote{This procedure is not strictly of one-loop order. It was shown, however, in Ref.~\cite{Frank:2003tg} 
for the MSSM that this procedure gives much exacter values for the Higgs mass including implicitly higher order corrections than a strict treatment at one-loop level. }
To calculate the one-loop mass of the $n^{th}$ Higgs boson 
the external momentum squared  $p^2$ in the renormalized self-energies
$\hat{\Sigma}_{ij}$ is  set equal to the
tree-level mass squared $(p^2=(M^{(0)}_{H_n})^2)$ in the first iteration step. Then, the mass matrix part of
$\hat{\Gamma}$, \textit{i.e.}~$\mathcal{M}^{\text{1l}}$, is
diagonalized. The thus obtained $n^{th}$ eigenvalue is the first approximation of the squared one-loop mass. In
the next iteration step $p^2$ is set equal to this value and once again the eigenvalues of $\mathcal{M}^{\text{1l}}$ are calculated to yield the next approximation of the one-loop mass. 
This iteration procedure is repeated until a precision of $10^{-9}$ is reached. 
All five Higgs boson masses are calculated this way.
\par
Note that in Eq.~\eqref{eq:sematrix} the mixing
with the Goldstone bosons is not taken into account but we have checked
numerically that the effect is negligible. 
Furthermore, it was shown in Ref.~\cite{Hollik:2002mv}, that in the MSSM it is sufficient to include the
 mixing with the Goldstone boson, whereas the mixing with the longitudinal component of the $Z$~boson does
 not have to be added explicitly. Taking into account the mixing of the Goldstone boson as well as the
mixing of the $Z$~boson leads to the same result as only including the Goldstone boson mixing.
\par
Due to the radiative corrections, not only the masses of the particles receive contributions but at the same time the tree-level mass eigenstates mix to form new one-loop mass eigenstates. 
In order to take this into account the Higgs mixing matrix ${\mathcal R}$ which performs the rotation from the interaction eigenstates to the mass eigenstates has to be adjusted so that
\begin{align}
H_i^{1l} = \mathcal R_{ij}^{1l} \Phi_j~.
\end{align}
In the numerical analysis, for the simplicity of the notation we drop the superscript $1l$ again. If not explicitly mentioned one-loop corrections are included.
 
The rotation of the tree-level to the one-loop mass eigenstates could be obtained by calculating finite wave function correction factors. The procedures to calculate these for a $2\times2$ or $3\times3$ mass matrix are described in Ref.~\cite{1lfull} and need to be extended for the $5\times5$ case. Another option is to apply the $p^2=0$ approximation. After setting the momenta in $\mathcal{M}^{\text{1l}}$ to zero, the rotation matrix that relates the tree-level to the one-loop mass eigenstates can be defined as the matrix that diagonalizes $\mathcal{M}^{\text{1l}}$.
The latter procedure has the advantage that the mixing matrix is unitary. The drawback is that it does not retain the full momentum dependence. For our numerical analysis we used the $p^2=0$ approximation for the determination of the mixing matrix. But we checked numerically that the differences between both methods are negligible.
\section{\label{sec:numerical} Numerical Analysis}
The calculation of the one-loop corrected Higgs boson masses has been
performed in two different calculations. While in one calculation
the Feynman rules have been derived from the NMSSM Lagrangian and
implemented in a {\tt FeynArts} model file \cite{feynarts}, they have
been obtained with the Mathematica package {\tt SARAH} \cite{sarah}
in the second
calculation and cross-checked against the first calculation. The
self-energies and tadpoles have been evaluated with the help of {\tt
  FormCalc} \cite{formcalc} in the 't Hooft-Feynman gauge. The divergent
integrals, regularized in the constrained differential renormalization
scheme \cite{constrained}, have been computed numerically with {\tt
  LoopTools} \cite{formcalc}. For the evaluation of the
counterterms, numerical diagonalization of the one-loop corrected
Higgs boson mass matrix and the determination of the mass
eigenvalues finally two independent Mathematica programs have been
written.  \s
 
We follow the SUSY Les Houches Accord (SLHA) \cite{slha} and compute
the parameters $M_W^2$ and $e$ of our input set defined in
Eq.~(\ref{eq:defparset}) 
from the SLHA pre-defined input values for the Fermi constant $G_F =
1.16637 \cdot 10^{-5}$~GeV$^{-2}$, the $Z$ boson mass $M_Z =
91.187$~GeV and the electroweak coupling $\alpha = 1/137$. If
not stated otherwise, we use the running $\overline{\mbox{DR}}$ top
quark mass $m_t$ at a common scale $Q=\sqrt{m_{Q_3} m_{t_R}}$. It is
obtained from the top quark pole mass $M_t=173.2$~GeV by taking the
routines of {\tt NMSSMTools} \cite{nmssmtools}.  
In the same way we obtain the running
$\overline{\mbox{DR}}$ bottom quark mass starting from the SLHA input
value $m_b (m_b)^{\overline{\scriptsize\mbox{MS}}}=4.19$~GeV. For the light
quarks we chose $m_u=2.5$~MeV, $m_c=1.27$~GeV, $m_d=4.95$~MeV, 
$m_s=101$~MeV \cite{pdg} and for the $\tau$ mass
$m_\tau=1.777$~GeV. \s 

In the following we exemplify the effects of complex phases in the
one-loop corrections to the NMSSM Higgs boson masses in different
scenarios. We require the scenarios to be compatible with
the recent results of the LHC Higgs boson searches
\cite{atlashiggs,cmshiggs} in the limit of the real NMSSM, {\it i.e.}
for vanishing CP-violating phases.\footnote{Note, that
  this is only an arbitrary choice which was made for practical
  reasons. We could as well have demanded a complex NMSSM scenario to be
  compatible with the recent LHC searches.} Our starting points
are the NMSSM benchmark points presented in Ref.~\cite{King:2012is}
which have been slightly modified for our analysis. With not too heavy
stop masses and not too substantial 
mixing they avoid unnaturally large finetuning, and
$\lambda$ and $\kappa$ have been chosen such that unitarity is not
violated below the GUT scale. Furthermore, we paid attention to keep
the effective $\mu \le 200$~GeV, with $\mu=\lambda\,v_s/\sqrt{2}$, in order not to violate tree-level
naturalness. For each scenario we verified that the non SM-like Higgs
bosons are not excluded by the searches at LEP \cite{LEPHbb}, Tevatron
\cite{tevsearch} and LHC. This 
has been cross-checked by running the program {\tt HiggsBounds}
\cite{higgsbounds}\footnote{The program {\tt NMSSMTools}
  \cite{nmssmtools} also performs these checks. It can be used, however,
  only for the case of the real NMSSM.}, which needs the complex
NMSSM Higgs couplings and branching ratios. The latter have been obtained by
adapting the Fortran code {\tt HDECAY} \cite{hdecay,susyhit} to the complex
NMSSM, in which we use the one-loop corrected Higgs boson masses
and mixing matrix elements of our calculation. For the SM-like Higgs boson
$H_i^{\mbox{\scriptsize SM-like}}$ of the real NMSSM we demand
its mass to lie in the interval $120-130$~GeV. Furthermore, the
total significance for $H_i^{\mbox{\scriptsize SM-like}}$ should not deviate by
more than 20\% from the corresponding SM value. We roughly estimate the
significance $S$ to be given by $S=N_s/\sqrt{N_b}$, where $N_s$ denotes the 
number of signal events and $N_b$ the number of background events. 
Hence our criteria for a scenario to be
compatible with present LHC searches are
\beq
120 \mbox{ GeV} &\le& M_{H_i^{\mbox{\scriptsize SM-like}}} \le 130 \mbox{
  GeV}
\\
S_{\scriptsize\mbox{tot}}^{\scriptsize\mbox{NMSSM}}
(H_i^{\mbox{\scriptsize 
    SM-like}}) &=&
S_{\scriptsize\mbox{tot}}^{\scriptsize\mbox{SM}} (H^{\mbox{\scriptsize 
    SM}}) \pm 20\%
\qquad 
\mbox{for} \qquad M_{H^{\mbox{\scriptsize 
    SM}}} =  M_{H_i^{\mbox{\scriptsize SM-like}}}\;.
\label{eq:sigincl}
\eeq
In this case the scenario is
estimated to be compatible with the present LHC searches taking into
account experimental and theoretical uncertainties. We roughly
approximate the total significance by adding in quadrature the
significances of the various LHC Higgs search channels. We assume the
number of background events to be the same both in the SM and the NMSSM
case. For the calculation of the signal events we need the cross section
values in the different channels, which we
obtain as follows. We first calculate the inclusive production cross
section by adding the gluon
fusion, weak boson fusion, Higgs-strahlung and $t\bar{t}$ Higgs
production cross sections. Associated production with $b\bar{b}$ does
not play a role here, as the $\tan\beta$ values we chose are rather
low. The gluon fusion value at NNLO QCD is obtained with {\tt
  HIGLU} \cite{higlu}, which we have modified to the NMSSM case. Weak
boson fusion and Higgs-strahlung at NLO QCD are computed with
the programs {\tt VV2H} and {\tt V2HV} \cite{programs} by applying the
modification 
factor due to the modified NMSSM Higgs coupling to gauge bosons
compared to the SM case. Finally, the cross section value for
$t\bar{t}$ Higgs production at NLO QCD \cite{nloqcdtth} is obtained
from the cross section values given at the LHC Higgs cross
section working group webpage \cite{higgswg} by applying the appropriate
factor taking into account the change of the NMSSM Higgs Yukawa
coupling with respect to the SM coupling. Note that the NLO QCD
corrections are not affected by changes due to the NMSSM Higgs sector
and can therefore readily be taken over from the SM case. The cross
sections in the $WW,ZZ$ and $\gamma\gamma$ LHC search channels are
obtained in the narrow width approximation by multiplication of the
total cross section with the corresponding Higgs branching ratios into
these final states. The branching ratios have been obtained from our 
modified Fortran code {\tt HDECAY} \cite{hdecay,susyhit}, adapted to the
complex NMSSM. The thus obtained cross sections for the various channels
can be used to calculate the number of signal events.\footnote{Note, that the
  luminosity factor in the calculation of the number of events and also
  the number of background events drop out in the comparison of the
  NMSSM case to the SM case, so that we only need to calculate the
  quadratic sum of the cross
  sections in the different final states for the NMSSM and for the SM and
  compare them.}
The experiments take into account QCD corrections
beyond NLO and also electroweak corrections. As these are not
available for the NMSSM we cannot take them into account here. They are of
the order of a few percent depending on the process. Furthermore,
ATLAS and CMS exploit more final states and combine them in a
sophisticated statistical procedure, while we have taken into
account only the most prominent ones. Our approximation should
therefore be viewed only as a rough estimate, good enough though to
eliminate scenarios clearly excluded by the present LHC search results. \s

In all investigated scenarios we have taken the input values at the
scale $Q=\sqrt{m_{Q_3} m_{t_R}}$. In order to comply with the present LHC
searches \cite{atlassquark,cmssquark}, we have throughout taken the soft
SUSY breaking mass parameters of the squarks of the 
first two generations equal to 1 TeV, and for simplicity also those of
the sleptons. The corresponding trilinear couplings are taken to be
1 TeV. Furthermore, the right-handed soft SUSY breaking mass parameter of the
sbottom sector is set equal to 1 TeV and its trilinear
coupling close to 1 TeV, so that we have
\beq
m_U &=& m_D \;=\; m_{Q_{1,2}}\; =\; m_E \;= \;m_L\; =\; 1 \; \mbox{TeV} \nonumber \\
A_x &=& 1 \; \mbox{TeV} \qquad (x=u,c,d,s,e,\mu,\tau)
\nonumber \\
A_b &\approx& 1 \; \mbox{TeV} \;.
\eeq
This leads to masses of $\sim 1$~TeV for the squarks of the
first and second family, the sleptons and the heavier
sbottom. Furthermore, all scenarios lead to the correct relic density in
the limit of the real NMSSM,
which has been checked with {\tt NMSSMTools} which contains a link  to
{\tt MicrOMEGAs} \cite{micromegas}.

\subsection{Scenario with a SM-like \boldmath{$H_3$}
  \label{sec:h3smlike}} 
The parameter set for this scenario is given by
\begin{align}
|\lambda|& = 0.72 \,, \quad |\kappa| = 0.20 \,, \quad \tan\beta = 3 \,, \quad
M_{H^\pm} = 629\mbox{ GeV} \,, \quad |A_\kappa| = 27\mbox{ GeV}
\,, \quad |\mu|= 198 \mbox{ GeV}\nonumber \\
|A_b| &= 963 \mbox{ GeV} \,,\quad 
|A_t| = 875 \mbox{ GeV} \,,  \quad M_1 = 145 \mbox{ GeV} \,, \quad M_2=
200 \mbox{ GeV} \,, \quad M_3= 600 \mbox{ GeV}\;.
\end{align}
The slightly high values of $\lambda$ and $\kappa$ may
require extra matter above the TeV scale
\cite{King:2012is}.\footnote{Being above the TeV scale it is not
  expected to influence LHC phenomenology, apart from the indirect
  effect of allowing $\lambda$ to be a somewhat larger than allowed by
  the usual perturbativity requirement in the NMSSM with no extra
  matter. 
  If instead one accepts more finetuning in
  the theory and allows for higher stop masses, a Higgs mass of the
  order of 125 GeV can be achieved for lower $\lambda$ values, {\it
    cf.}~the discussion in Sect.~\ref{sec:h1h2scenario}.} 
We set all CP-violating phases to zero and subsequently turn on specific
phases to study their respective influence. In this case, the signs of
the tree-level CP-violating phases Eqs.~(\ref{eq:cphixsign}),
(\ref{eq:cphizsign}) are then chosen as
\beq
\mbox{sign } \cos\varphi_x = +1 \;, \qquad
\mbox{sign } \cos\varphi_z = -1 \;. 
\eeq
Furthermore, the left- and right-handed soft SUSY breaking
mass parameters in the stop sector are given by $m_{Q_3}=490$~GeV and
$m_{t_R}=477$~GeV. This leads to relatively light stop masses
$m_{\tilde{t}_1}=363$~GeV and $m_{\tilde{t}_2}=616$~GeV, still allowed
by the experiments \cite{thirdgenatlas,thirdgencms}.\footnote{Note,
  that light stop masses and small mixing reduce the amount of finetuning
  \cite{King:2012is}.} 
In the calculation of the one-loop correction to the Higgs boson
masses we have set the renormalization scale equal to 500~GeV, {\it
  i.e.}~$\mu_{ren}=500$~GeV, if not stated otherwise. This scenario
leads in the CP-conserving NMSSM to the one-loop corrected $H_3$ being
SM-like with a mass $M_{H_3}= 125$~GeV compatible with present LHC
searches. In the following we discuss for various complex phase
choices the phenomenology of the three lightest Higgs bosons. The two
heavier ones receive mass corrections of maximally 2~GeV leading to
masses of $\sim 642$~GeV so that they are not excluded by
present collider searches, with $H_4$ being mostly CP-odd and
$H_5$ mostly CP-even. We therefore do not display their masses
explicitly.

\subsubsection{CP violation at tree-level} 
\begin{figure}[b]
\begin{center}
\includegraphics[width=0.49\linewidth]{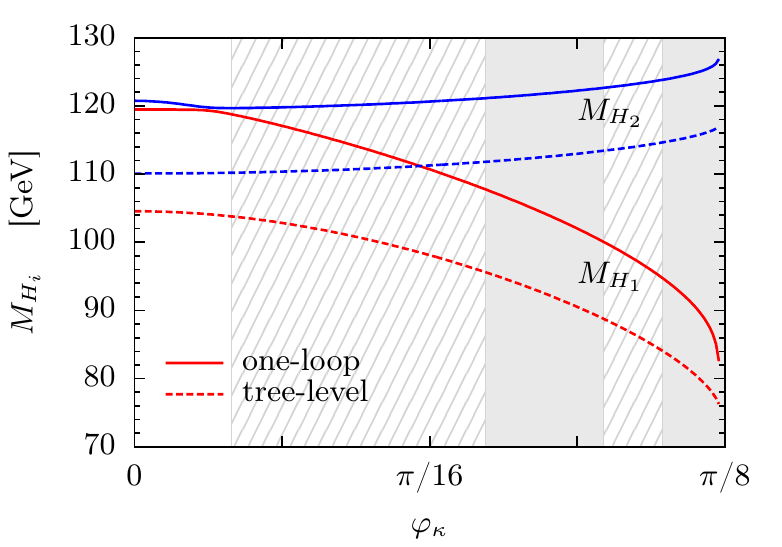}
\includegraphics[width=0.49\linewidth]{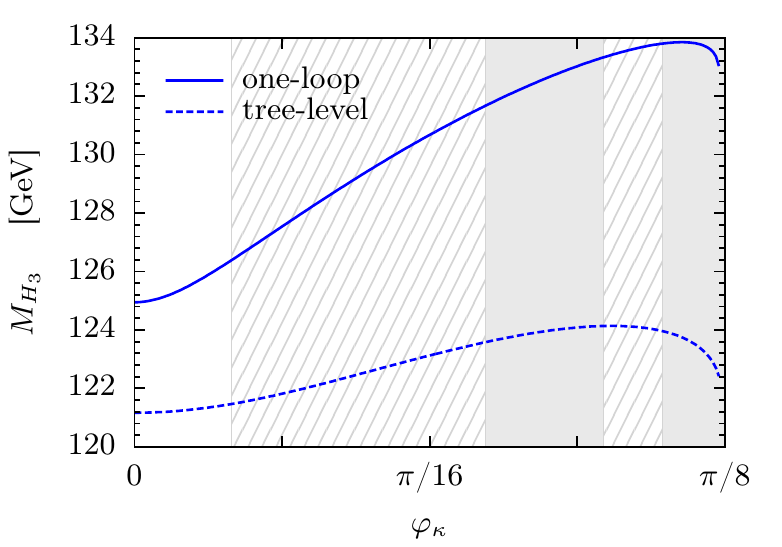}
\end{center}
\vspace*{-0.5cm}
\caption{Left: Tree-level (dashed) and one-loop corrected (full) Higgs
  boson masses for $H_1$ (red) and $H_2$ (blue)
  as a function of $\varphi_\kappa$. Right: Tree-level (dashed) and
  one-loop (full) mass $M_{H_3}$ as a function of $\varphi_\kappa$. The
  exclusion region due to LEP, Tevatron and LHC data is shown as grey
  area, the region with the SM-like Higgs boson not being compatible
  with an excess of data around 125 GeV as dashed area.}
\label{fig:all3varphikappa}
\end{figure}
As we have seen in Sect.~\ref{sec:tree} a non-vanishing phase $\varphi_y$,
{\it cf.}~Eq.~(\ref{eq:varphiy}), introduces CP violation at
tree-level. Therefore CP-even and CP-odd Higgs mass eigenstates cannot
be distinguished any more. A measure for CP violation
concerning the state $H_i$ ($i=1,...,5$) is
instead provided by the quantity
\beq
r^i_{\scriptsize \mbox{CP}} \equiv ({\cal R}_{i1})^2 + ({\cal R}_{i2})^2 +
({\cal R}_{i3})^2 \; ,
\eeq
where ${\cal R}_{ij}$ are the matrix elements of the mixing matrix
which diagonalizes the Higgs boson mass matrix, {\it
  cf.}~Eq.~(\ref{eq:rndef}). A purely CP-even (CP-odd) mass eigenstate
$H_i$ corresponds to $r^i_{\scriptsize \mbox{CP}}=1$ (0). We first
investigate the effect of a non-vanishing phase\footnote{The choice of
  non-vanishing $\varphi_\kappa$ allows to 
  investigate mixing effects of the Higgs bosons while suppressing the
  phase relevant for the neutral electric dipole moment
  \cite{Funakubo:2004ka}.} 
\beq
\varphi_\kappa\ne 0 \;. 
\eeq
The phases $\varphi_{A_\lambda},\varphi_{A_\kappa}$ are fixed by the
tadpole conditions Eqs.~\eqref{tadad}, \eqref{tadas}. \s

\begin{figure}[t]
\begin{center}
\includegraphics[width=0.49\linewidth]{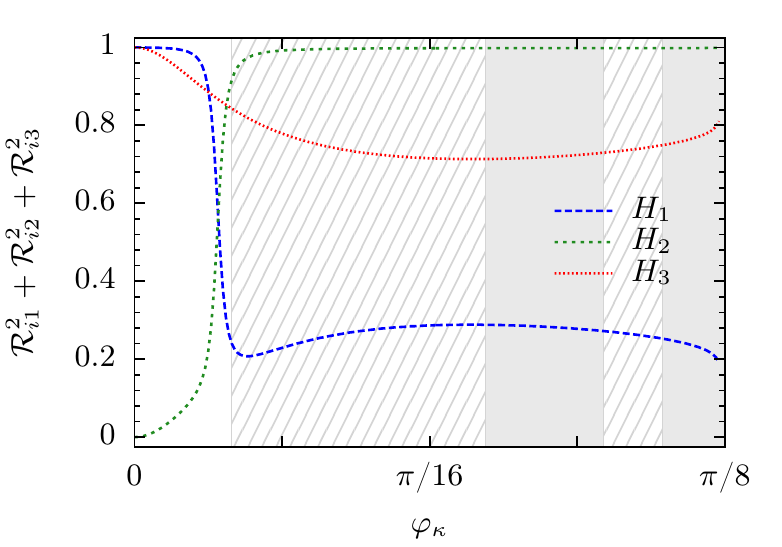}
\includegraphics[width=0.49\linewidth]{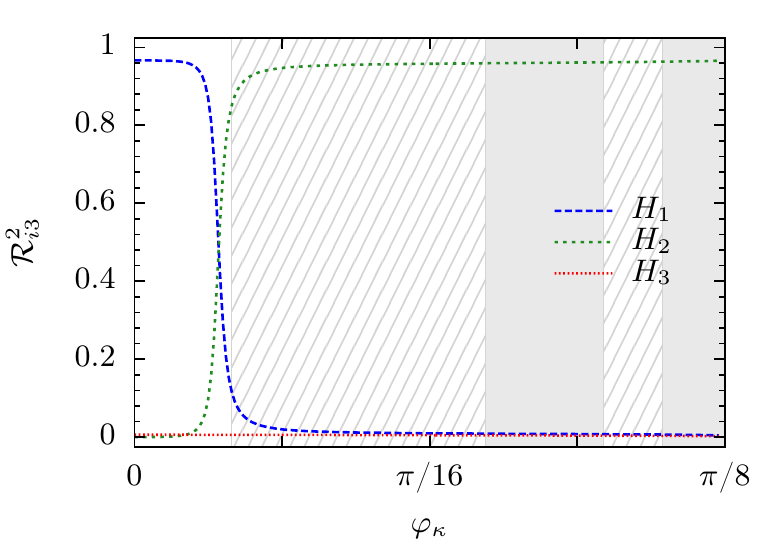}
\end{center}
\vspace*{-0.5cm}
\caption{The amount of CP violation $r^i_{\scriptsize \mbox{CP}}$ for $H_i$ ($i=1,2,3$)
 as a function of $\varphi_\kappa$ (left). The amount of CP-even singlet
 component $({\cal R}_{i3})^2$ as a function of
 $\varphi_\kappa$ (right).} 
\label{fig:rivarphikappa}
\end{figure}
\begin{figure}[b]
\begin{center}
\includegraphics[width=0.49\linewidth]{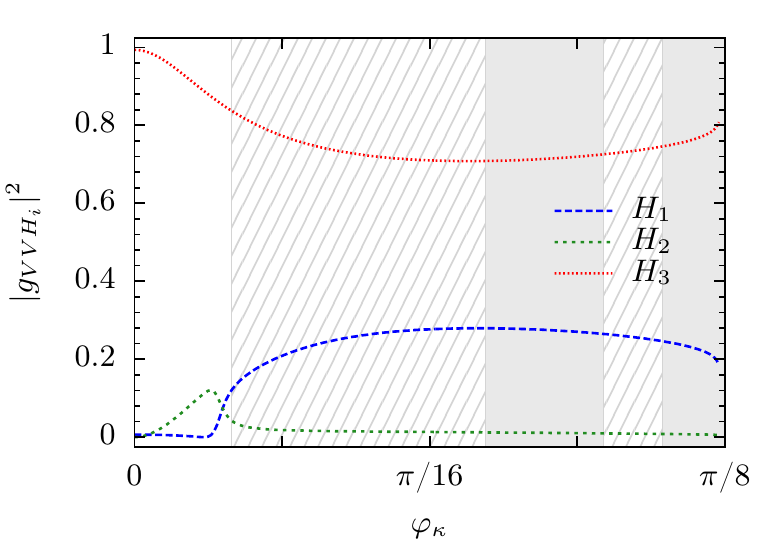}
\end{center}
\vspace*{-0.5cm}
\caption{The $H_i$ coupling to $V$ ($V=Z,W$)
  bosons squared ($i=1,2,3$) normalized to the SM coupling,
  $|g_{VVH_i}|^2$, as a function of $\varphi_\kappa$.} 
\label{fig:zcoupvarphikappa}
\end{figure}
In Fig.~\ref{fig:all3varphikappa} (left) we show the tree-level and
one-loop masses of the two lightest Higgs mass eigenstates
$H_{1,2}$ as a function of $\varphi_\kappa$, where $\varphi_\kappa=0$
corresponds to the real NMSSM. The phase is varied up to
$\pi/8$. Above this value it turns out that the phases $\varphi_{A_{\lambda}}$ and $\varphi_{A_{\kappa}}$ cannot be chosen in such a way that the tadpole conditions
are fulfilled. The tree-level and one-loop corrected
mass of the SM-like $H_3$ is shown in
Fig.~\ref{fig:all3varphikappa} (right). Figure~\ref{fig:rivarphikappa}
displays, as a function of $\varphi_\kappa$, the amount of CP
violation $r^i_{\scriptsize \mbox{CP}}$ (left) and the amount of the
CP-even singlet component\footnote{The CP-odd singlet component is given
by $({\cal R}_{i5})^2$.} $({\cal R}_{i3})^2$ (right) for $H_{1,2,3}$. Finally 
Fig.~\ref{fig:zcoupvarphikappa} shows their coupling squared to
the $V$ bosons ($V=Z,W$) normalized to the SM as a function of
$\varphi_\kappa$. As expected, the masses exhibit already at
tree-level a sensitivity to the CP-violating phase $\varphi_\kappa$. In
particular for the SM-like $H_3$ this dependence is more pronounced at
one-loop level, changing its mass value by up to 9 GeV for $\varphi_\kappa \in
[0,\pi/8]$. The one-loop correction increases the mass by $\sim 4$
to 11~GeV depending on $\varphi_\kappa$ with larger mass values for
larger CP-violating phases. \s

In the plots the grey areas are the parameter regions which are
excluded due to the experimental constraints from LEP, Tevatron and
LHC, and which have been obtained with {\tt HiggsBounds}.\footnote{The
  exclusion is due to the LEP constraint on $H_1$ from the Higgs boson search in
  the $Zb\bar{b}$ final state stemming from a Higgs boson produced in
  Higgs-strahlung with subsequent decay into a $b$-quark pair.} 
This is the case for $0.074 \, \pi < \varphi_\kappa < 0.099 \, \pi$
and $\varphi_\kappa > 0.112 \, \pi$. The dashed region excludes the
parameter regions 
where the criteria stated in Eq.~(\ref{eq:sigincl}) of compatibility with the
recent Higgs excess around 125 GeV cannot be fulfilled any more, here
for $\varphi_\kappa > 0.021 \pi$. The reason is that with increasing
$\varphi_\kappa$ the eigenstate $H_3$ becomes more CP-odd and hence couples
less to $VV$ ($V=Z,W$) as can be inferred from Fig.~\ref{fig:rivarphikappa}
(left) and Fig.~\ref{fig:zcoupvarphikappa} so that the total cross
section becomes smaller and the significance deviates by more than 20\%
from the SM significance of a SM Higgs boson with same mass. \s

The one-loop corrections for the two lighter Higgs bosons $H_{1,2}$
increase their masses by 6-15~GeV depending on $\varphi_\kappa$. The
mass value $M_{H_1(H_2)}$ decreases (increases) with rising
$\varphi_\kappa$. In
the CP-conserving limit the one-loop masses of $H_{1,2}$ are
$M_{H_1}=119.4$~GeV and $M_{H_2}=120.7$~GeV, with $H_1$ being CP-even, {\it
  cf.}~Fig.~\ref{fig:rivarphikappa} (left), but CP-even singlet-like, {\it
  cf.}~Fig.~\ref{fig:rivarphikappa} (right), such that it hardly
couples to SM particles and cannot be excluded by the experimental
searches. The heavier Higgs $H_2$ is dominantly CP-odd singlet-like
(not plotted here) and is not
excluded by the LEP, Tevatron and LHC searches due to both its singlet and its
CP-odd nature leading to a vanishing coupling to weak vector bosons, {\it cf.}
Fig.~\ref{fig:zcoupvarphikappa}. With increasing CP-violating phase the
eigenstates $H_1$ and $H_2$ interchange their roles both with respect
to their CP nature and their amount of CP-even singlet component, with the
cross-over taking place at $\varphi_\kappa \approx \pi/64$. \s

\begin{figure}[t]
\begin{center}
\includegraphics[width=0.49\linewidth]{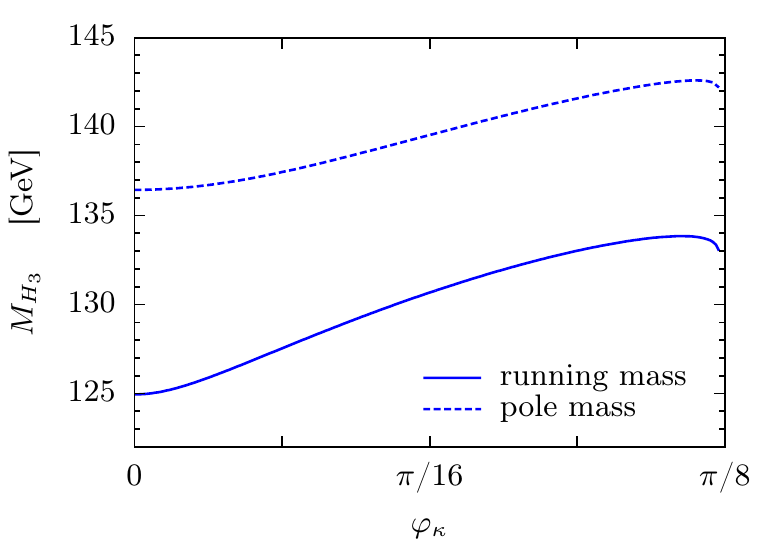}
\end{center}
\vspace*{-0.5cm}
\caption{The one-loop corrected mass of the SM-like Higgs $H_3$
  evaluated with the top and bottom running $\overline{\mbox{DR}}$
  masses (full) and  with the corresponding pole masses (dashed).} 
\label{fig:schemecompar}
\end{figure}
In order to get an estimate of the theoretical uncertainty due to the
unknown higher-order corrections the one-loop corrections to the Higgs
boson masses have been calculated with the top and bottom pole quark
masses, $M_{t}=173.2$~GeV and $M_b=4.88$~GeV, and compared to
the results for the one-loop corrected masses 
evaluated with the running $\overline{\mbox{DR}}$ top and bottom quark
masses $m_{t,b}$ at the scale $Q=\sqrt{m_{Q_3}m_{t_R}}$. For our
scenario they amount to 
$m_t = 153.4$~GeV and $m_b=2.55$~GeV. The result is shown in
Fig.~\ref{fig:schemecompar}. Whereas the slope of the curve hardly
changes, the absolute values of the corrections change and are more
important for a higher top quark 
mass. The theoretical uncertainty due to the different quark mass
renormalization schemes can conservatively be estimated to $\sim 10$\%.

\subsubsection{No tree-level CP violation}
\begin{figure}[b]
\begin{center}
\includegraphics[width=0.49\linewidth]{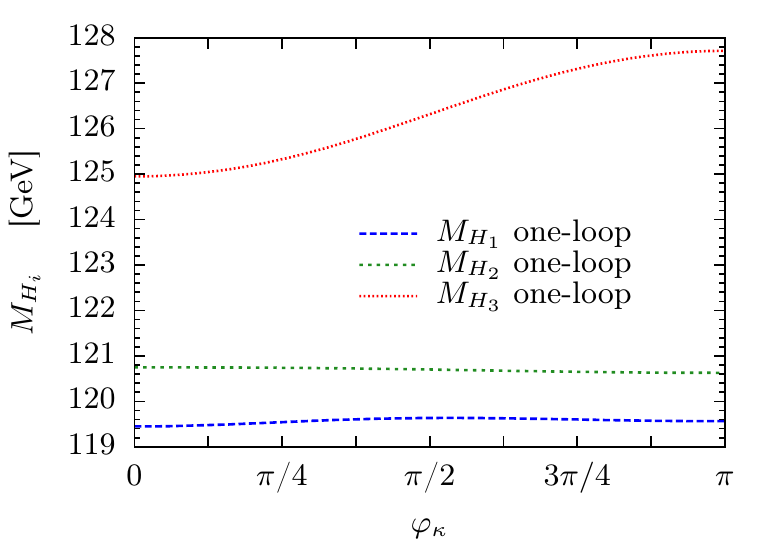}
\includegraphics[width=0.49\linewidth]{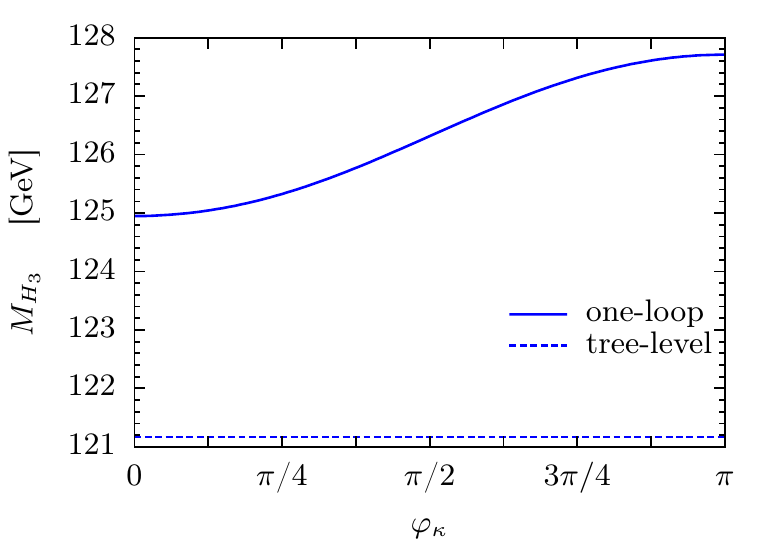}
\end{center}
\vspace*{-0.5cm}
\caption{One-loop corrected Higgs boson masses $M_{H_i}$ ($i=1,2,3$)
  as a function of $\varphi_\kappa=\varphi_\lambda$ (left). Tree-level
  (dashed) and one-loop corrected (full) mass for $H_3$ as a function of
  $\varphi_\kappa=\varphi_\lambda$ (right).}
\label{fig:all3notreecp}
\end{figure}

\begin{figure}[ht]
\begin{center}
\includegraphics[width=0.49\linewidth]{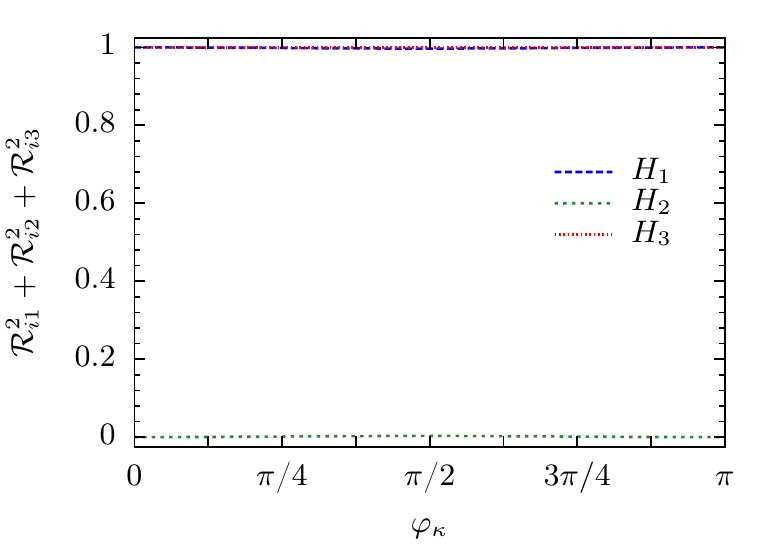}
\includegraphics[width=0.49\linewidth]{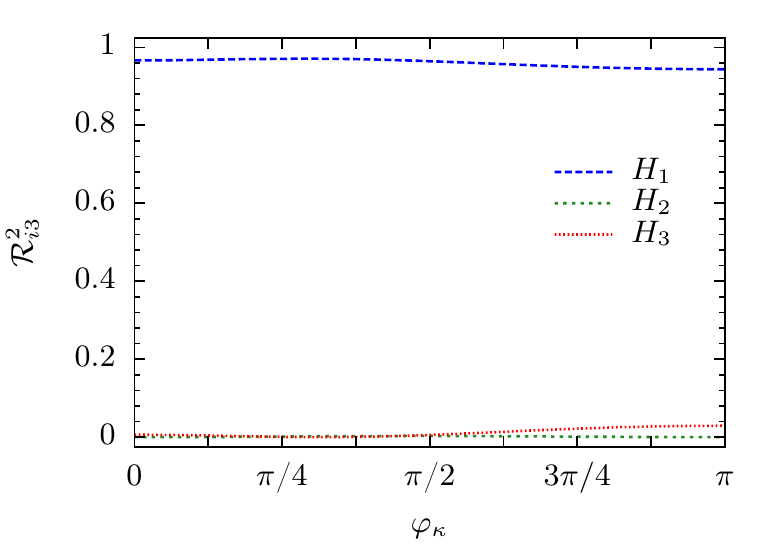}
\end{center}
\vspace*{-0.5cm}
\caption{The amount of CP violation $r^i_{\scriptsize \mbox{CP}}$ for $H_i$ ($i=1,2,3$)
  as a function of $\varphi_\kappa=\varphi_\lambda$ (left). The amount
  of CP-even singlet component $({\cal R}_{i3})^2$ as a function of
  $\varphi_\kappa=\varphi_\lambda$ (right).} 
\label{fig:rcp}
\end{figure}

We now keep the CP-violating phases $\varphi_\kappa$ and
$\varphi_\lambda$ non-zero and vary them by the same amount, such that 
according to Eq.~(\ref{eq:varphiy}) we have no tree-level CP-violating
phase $\varphi_y$. In the one-loop corrections
$\varphi_\kappa,\varphi_\lambda$ enter separately so that CP violation
is induced radiatively. 
Figure~\ref{fig:all3notreecp} (left) shows the
one-loop corrected masses of the three lightest Higgs states
$H_{1,2,3}$, Fig.~\ref{fig:all3notreecp} (right) compares the
tree-level and one-loop corrected mass of the SM-like $H_3$, both as a function of
$\varphi_\kappa$. The tree-level mass shows no dependence on
$\varphi_\kappa$ as expected. 
The one-loop mass $M_{H_3}$
changes by only $\sim 3$~GeV for $\varphi_\kappa$ varying from 0 to
$\pi$, and the loop-corrected masses for $H_{1,2}$ show almost no
dependence on the CP-violating phase. The reason is that the
dependence on the phase is due to the corrections from the stop sector
which are the dominant contributions to the one-loop masses. The
values of the stop masses change with the CP-violating phase. As $H_3$
has the largest $h_u$ component and hence couples more strongly to the
up-type quarks it shows a stronger dependence on
$\varphi_\kappa$ than $H_1$ and $H_2$.
For $M_{H_1}$ ($M_{H_2}$) the mass corrections are of about 15
(11)~GeV. With mass values around 120 GeV they could lead to
additional signals at the LHC if they were SM-like. However, due to
the CP-odd nature of 
$H_2$, {\it cf.}~Fig.~\ref{fig:rcp} (left), it hardly couples to weak
vector bosons. And the CP-even singlet character of $H_1$, compare with
Fig.~\ref{fig:rcp} (right), reduces its couplings to SM
particles. These particles would therefore have considerably reduced
signals at the LHC. The whole region over which
$\varphi_\kappa=\varphi_\lambda$ are varied is hence still allowed by
the LHC searches.

\subsubsection{Radiatively induced CP violation through the stop sector}
\begin{figure}[h]
\begin{center}
\includegraphics[width=0.49\linewidth]{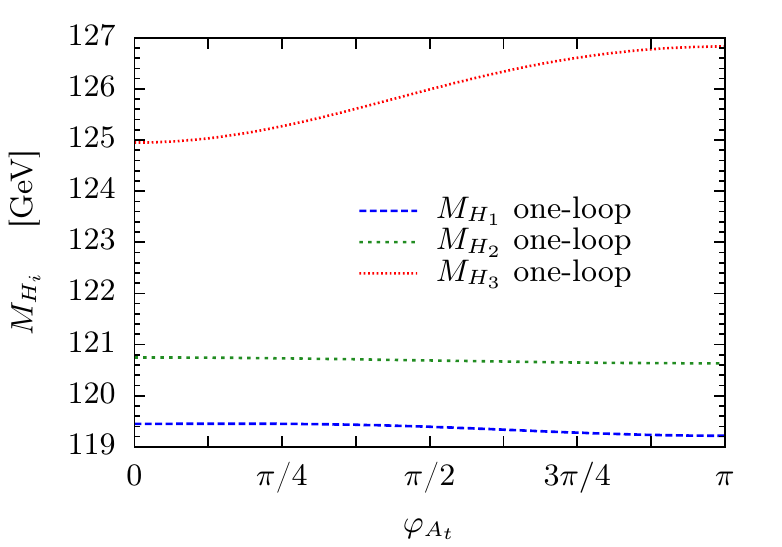}
\includegraphics[width=0.49\linewidth]{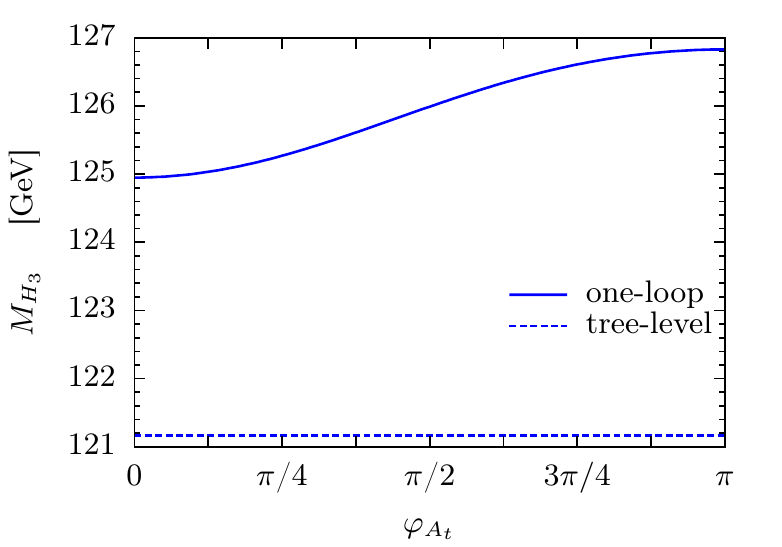}
\end{center}
\vspace*{-0.5cm}
\caption{One-loop corrected Higgs boson masses $M_{H_i}$ ($i=1,2,3$)
  as a function of $\varphi_{A_t}$ (left). Tree-level (dashed) and
  one-loop (full) mass $M_{H_3}$ as a function of $\varphi_{A_t}$ (right).} 
\label{fig:atvar}
\end{figure}

For completeness we investigate the case where only $\varphi_{A_t} \ne
0$. CP violation is thus only induced 
through loop corrections stemming from the stop sector. The one-loop
corrected masses of $H_{1,2,3}$ are shown in Fig.~\ref{fig:atvar}
(left), the tree-level and one-loop corrected mass of the SM-like
$H_3$ are displayed separately in Fig.~\ref{fig:atvar} (right), both as
function of $\varphi_{A_t}$. The tree-level masses of $H_{1,2}$ are
increased by about 10-15~GeV and their one-loop masses of $\sim
119.5$ and 121~GeV, respectively, hardly show any dependence on
$\varphi_{A_t}$. 
The SM-like $H_3$ one-loop mass shows a small
dependence varying by $\sim 2$~GeV 
for $\varphi_{A_t} \in 
[0,\pi]$, increasing the tree-level mass by 4-7~GeV. 
The reason is that $H_3$ has the
largest $h_u$ component so that the dominant one-loop corrections
stemming from the stop loops contribute more importantly to the
radiative corrections of the Higgs mass matrix elements of
$H_3$ than of $H_1$ and $H_2$.
The CP-even singlet
nature of $H_1$, the CP-even character of $H_1$ and $H_3$ as well as the CP-odd one
of $H_2$ are hardly affected by a change in $\varphi_{A_t}$ and
are therefore not displayed here. As may have been expected, in this scenario
loop-induced CP violation 
affects the phenomenology of the Higgs bosons less. Note, that the
scenario is not excluded by LHC searches over the whole displayed
phase range. \s

\begin{figure}[h]
\begin{center}
\includegraphics[width=0.49\linewidth]{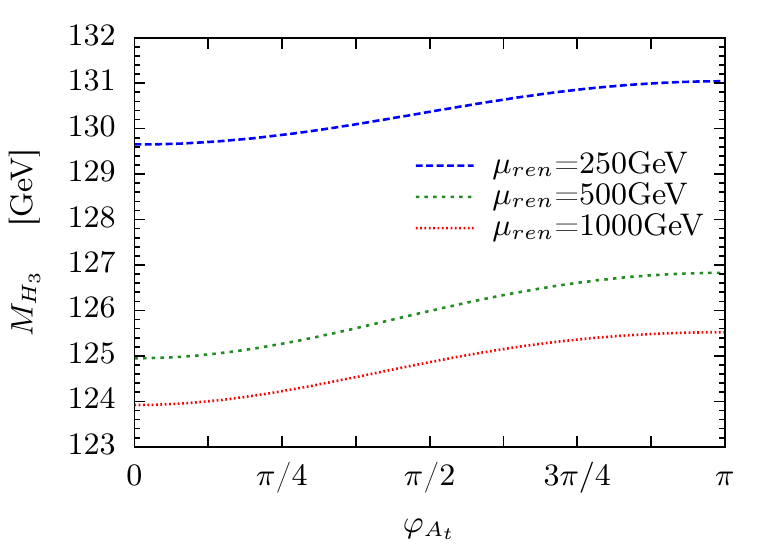}
\end{center}
\vspace*{-0.5cm}
\caption{One-loop corrected Higgs boson masses $M_{H_3}$ as a function
  of $\varphi_{A_t}$ for three different renormalization scales,
  $\mu_{ren}=250$ (blue/long-dashed), $500$ (green/short-dashed) and 1000~GeV
  (red/dotted).} 
\label{fig:scalevar}
\end{figure}

In Fig.~\ref{fig:scalevar} we investigate the theoretical error due to
unknown higher-order corrections by varying the renormalization scale
from 500 GeV to half and twice the scale. The variation of the
renormalization scale also changes the values of the input parameters
and the running
$\overline{\mbox{DR}}$ top and bottom mass. For higher scales they
become smaller and hence also the one-loop corrections to the
masses decrease. 
The residual theoretical uncertainty can be estimated to about $4$\%.
We also checked the theoretical uncertainty due to the different quark mass
renormalization schemes and found them to be of $\sim 10$\%, hence of
the same order as in the scenario studied in Sect.~\ref{sec:h3smlike}.

In summary, the discussion of the various scenarios has shown that
the impact of the CP-violating phase is
crucial for the validity of the model. While a certain parameter set can
still accommodate the experimental results for vanishing CP violation it
may be invalidated by non-vanishing CP phases.
Turning this around, the
experimental results will be useful to pin down the allowed amount of CP
violation. The latter can arise from tree-level CP-violating phases in
the Higgs sector or be radiatively induced. In the latter case the
effects are found to be less pronounced.
To get reliable predictions, the one-loop corrections have to be
included as they not only considerably change the absolute mass values but also
the singlet and CP-nature of the individual Higgs bosons as compared to
the tree-level quantities.
\subsection{Scenario with SM-like \boldmath{$H_1$} or
  \boldmath{$H_2$} \label{sec:h1h2scenario}}
The parameter set for this scenario, where, depending on the
CP-violating phase, either $H_1$ or $H_2$ is SM-like,
is given by
\begin{align}
|\lambda| &= 0.65 \,, \quad |\kappa| = 0.25 \,, \quad \tan\beta = 3 \,, \quad
M_{H^\pm} = 619\mbox{ GeV} \,, \quad |A_\kappa| = 18\mbox{ GeV}
\,, \quad |\mu|= 199 \mbox{ GeV} \nonumber \\
|A_b| &= 971 \mbox{ GeV} \,,\quad |A_t| = 1143 \mbox{ GeV} \,,  \quad M_1 = 105 \mbox{ GeV} \,, \quad M_2=
200 \mbox{ GeV} \,, \quad M_3= 600 \mbox{ GeV}\;.
\end{align}
The signs of the tree-level CP-violating phases
Eqs.~(\ref{eq:cphixsign}) and (\ref{eq:cphizsign}) are
\beq
\mbox{sign } \cos\varphi_x = +1 \; , \qquad
\mbox{sign } \cos\varphi_z = -1 \;. 
\eeq
The renormalization scale has been set to $\mu_{ren}=650$~GeV.
The left- and right-handed soft SUSY breaking
mass parameters in the stop sector  $m_{Q_3}=642$~GeV and $m_{t_R}= 632$~GeV
lead to $m_{\tilde{t}_1}=514$~GeV and $m_{\tilde{t}_2}=768$~GeV. The
low value of $\lambda$ respects the
bounds imposed by unitarity \cite{King:2012is}. We allow for
tree-level CP violation by choosing $\varphi_\kappa \ne 0$. The
remaining complex phases are all set to zero, except for
$\varphi_{A_\lambda}$ and $\varphi_{A_\kappa}$ which follow from the
tadpole conditions Eqs.~\eqref{tadad}, \eqref{tadas}. \s

\begin{figure}[b]
\begin{center}
\includegraphics[width=0.49\linewidth]{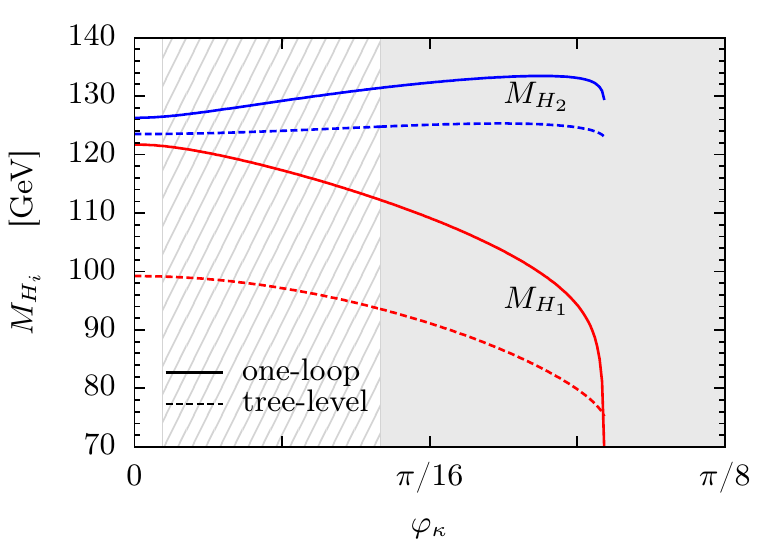}
\includegraphics[width=0.49\linewidth]{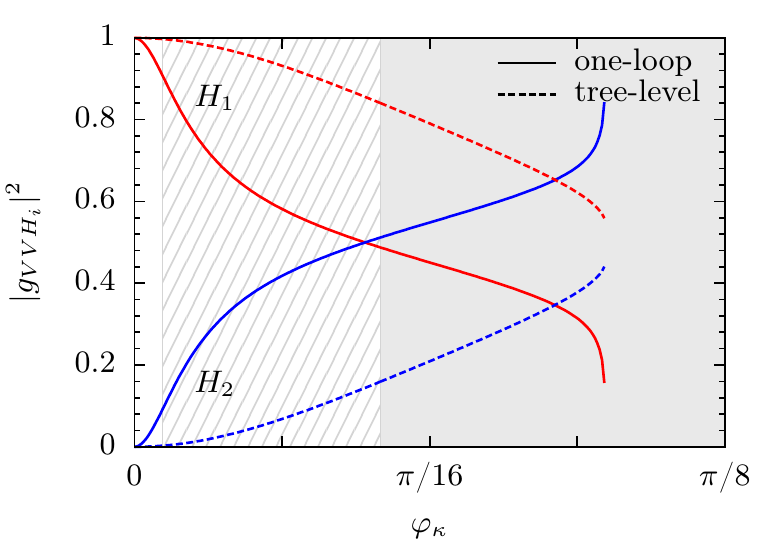}
\end{center}
\vspace*{-0.5cm}
\caption{Left: Tree-level (dashed) and one-loop corrected (full) Higgs
  boson masses as a function of $\varphi_\kappa$ for $H_1$ (red) and
  $H_2$ (blue). Right: The $H_{1}$ (red) and $H_2$ (blue) Higgs
  couplings squared to two $V$ bosons ($V=W, Z$) as a function of $\varphi_\kappa$ at
  tree-level (dashed) and at one-loop (full). The
  exclusion region due to LEP, Tevatron and LHC data is shown as grey
  area, the region with the SM-like Higgs boson not being compatible
  with an excess of data around 125 GeV as dashed area.}
\label{fig:nmp1mass12}
\end{figure}
\begin{figure}[b]
\begin{center}
\includegraphics[width=0.49\linewidth]{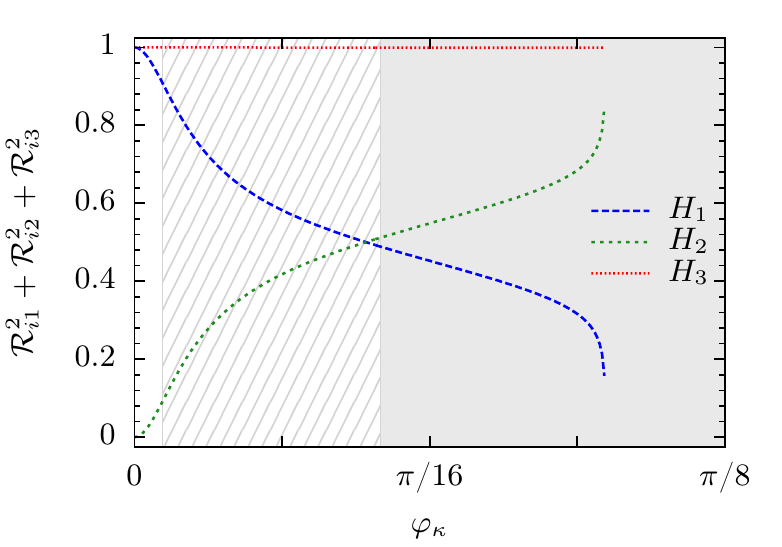}
\includegraphics[width=0.49\linewidth]{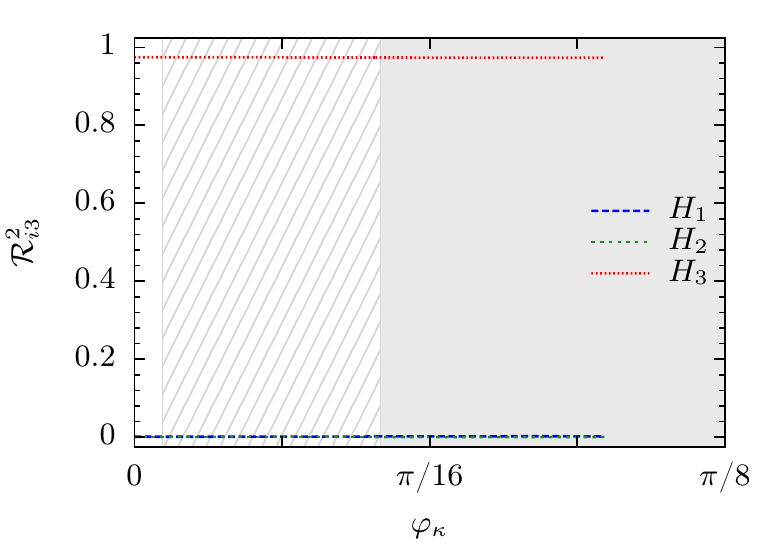}
\end{center}
\vspace*{-0.5cm}
\caption{The amount of CP violation $r^i_{\scriptsize \mbox{CP}}$ for $H_i$ ($i=1,2,3$)
  as a function of $\varphi_\kappa$ (left). The amount of CP-even singlet
  component $({\cal R}_{i3})^2$ as a function of
  $\varphi_\kappa$ (right).} 
\label{fig:nmp1char}
\end{figure}
The tree-level and and one-loop corrected masses of $H_1$ and $H_2$
are shown in Fig.~\ref{fig:nmp1mass12} (left), as a function of
$\varphi_\kappa$. Beyond
$\varphi_\kappa \approx 0.1 \pi$ the tadpole conditions are not fulfilled any
more. The corresponding couplings squared to weak vector bosons are plotted in
Fig.~\ref{fig:nmp1mass12} (right). 
The amount of CP violation of the three lightest Higgs bosons
$H_{1,2,3}$ and their
CP-even singlet component are displayed in Fig.~\ref{fig:nmp1char} (left) and
(right), respectively, as a function of $\varphi_\kappa$. 
As can be inferred from the Figures, in the limit of the real NMSSM
$H_1$ is CP-even and has SM-like couplings while $H_2$ is CP-odd. \
The heavier $H_3$ is CP-even over the whole $\varphi_\kappa$ range. 
The CP-even singlet components of $H_1$ and $H_2$ vanish, {\it cf.}
Fig.~\ref{fig:nmp1char} (right). However, $H_2$ is CP-odd singlet-like. This is
reflected in the couplings of $H_1$ and $H_2$ to the weak vector bosons. The one-loop
corrections for $\varphi_\kappa=0$ shift the $H_1$ mass from 99 to about
122~GeV so that its mass is compatible with the excess observed at
the LHC. The mass of the second Higgs boson $H_2$ is increased by about 3 GeV to 126~GeV
and could have been observed at the LHC, if its coupling to gauge bosons were not suppressed due to its
CP-odd character so that it is not excluded by the present experimental
constraints from the LHC. Furthermore, for $\varphi_\kappa=0$ the
total significance concerning $H_1$ is compatible 
with LHC searches according to our criteria Eq.~(\ref{eq:sigincl}). \s

The scenario is interesting because with increasing $\varphi_\kappa$ the
CP character of $H_{1,2}$ changes rapidly (with a cross-over at $\varphi_\kappa \approx
3\pi/64$ where $H_{1,2}$ interchange their roles with $H_1$ being more
CP-odd like and $H_2$ more CP-even like). This dependence on the CP
violating phase is at one-loop more pronounced than at tree-level and
makes that already beyond
$\varphi_\kappa \approx 0.006 \, \pi$ $H_1$ cannot fulfill the role of the
SM-like Higgs boson any more as its couplings deviate too much from the SM
case to fulfill the requirement of Eq.~(\ref{eq:sigincl}). On the other hand,
the $H_2$ couplings are not yet SM-like and once this is the case $H_2$
is already too heavy to be compatible with LHC searches. Hence, above
$\varphi_\kappa \approx 0.006 \, \pi$ the scenario does not comply with the criteria of
Eq.~(\ref{eq:sigincl}) any more and it is excluded as indicated
by the dashed region. Beyond $\varphi_\kappa \approx 0.052 \, \pi$ the grey region
shows that the searches at LEP invalidate this parameter
choice due to the LEP limit on $H_1$ in $Zb\bar{b}$. Therefore a 
large portion of this scenario is likely to be excluded, constraining
$\varphi_\kappa$ to be almost zero and hence a real NMSSM. \s

This scenario illustrates particularly well the importance of the one-loop
corrections and the impact on the restriction of a possible CP-violating
phase. Firstly, the one-loop corrections are crucial to shift the mass
of the SM-like Higgs boson to a mass value which is compatible with
the excess observed at the LHC. However, the one-loop corrections also
amplify the dependence on the CP-violating phase of both the Higgs
masses and in particular the mixing matrix elements and hence the
coupling to the weak vector bosons. Neglecting for the moment 
for the sake of this discussion the
fact that at tree-level $H_1$ does not fulfill the mass constraint, the
restriction of the CP-violating phase due to deviations from the
SM significance would be less severe at tree-level than at
one-loop level due to the smooth tree-level dependence on the
CP-violating phase. The one-loop corrections are hence crucial to
correctly define parameter scenarios which are compatible with present
LHC searches and to derive the correct exclusion limits for scenarios
dropping out of this constraint. \s

We close this subsection with the discussion of the CP-even $H_3$, not
shown explicitly in all plots. It is
dominantly CP-even singlet-like. Its tree-level mass of $\sim 148-152$~GeV
for $\varphi_\kappa$ increasing from 0 to $\sim 3\pi/16$ receives
one-loop corrections of $\sim 3-4$~GeV. The one-loop corrections show
the same dependence on $\varphi_\kappa$ as the tree-level mass. Due to
its singlet character at present it cannot be excluded by LHC searches. \s

\section{\label{sec:summary} Summary and Conclusions} 
We have calculated the one-loop corrections to the neutral Higgs
bosons in the CP-violating NMSSM by applying a mixed
renormalization scheme where part of the parameters are renormalized
on-shell while $\tan\beta,v_s,\lambda,\kappa,A_\kappa$ and the CP-violating phases are renormalized in the $\overline{\mbox{DR}}$ scheme.
We have in general allowed for tree-level CP violation due to non-vanishing
phases $\varphi_u,\varphi_s,\varphi_\kappa$ and $\varphi_\lambda$, and for
loop induced CP 
violation from the stop sector due to a non-zero phase
$\varphi_{A_t}$. Several scenarios have been investigated which start
from parameter sets that are compatible with the experimental Higgs
searches in the limit of the real NMSSM, subsequently CP violation is
turned on. 
As expected the dependence of the one-loop corrected
Higgs masses and mixing matrix elements on the CP-violating phase
turned out to be 
more pronounced for tree-level CP violation than for radiatively
induced CP violation. The loop corrections were found to
considerably change the masses and mixing angles with crucial
implications for the Higgs phenomenology at the LHC. As it is well known in the MSSM and the real NMSSM we also found that a scenario
may be excluded at 
tree-level, whereas it is compatible with LHC searches at one-loop. Of special
interest is the dependence on the CP-violating phase. It may be rather
smooth at tree-level but more pronounced at one-loop so that at
one-loop the CP-violating phase under investigation may be much more
restricted than at tree-level due to possible non-compatibility with the experiments. Therefore, in order to correctly define viable
scenarios and pin down allowed parameter ranges the one-loop
corrections are indispensable. We also investigated the theoretical
error due to the unknown higher order corrections by applying an
on-shell and a $\overline{\mbox{DR}}$ renormalization scheme for the
top and bottom quark mass and by varying the renormalization scale
between half and twice its value. The theoretical error of the
one-loop corrected Higgs masses can be conservatively estimated to be about 10\%.

\section*{Acknowledgments}

\noindent
We would like to thank J.~Baglio, M.~Maniatis, M.~Spira and D.~St\"ockinger for helpful discussions.
This research was supported in part by the Deutsche
Forschungsgemeinschaft via the Sonderforschungsbereich/Transregio
SFB/TR-9 Computational Particle Phy\-sics. R.G. acknowledges financial
support from the Landesgra\-du\-ier\-tenkol\-leg.

\begin{appendix}
\section{\label{sec:trafo} Relations between Original and Physical Parameters}
For the transformation of the Lagrangian from the original parameters to the physical 
ones the following relations are used:
\begin{align}\label{Ersetzung1}
m_{H_d}^2 &= \frac{e}{2 c_\beta M_W s_{\theta_W}} t_{h_d} - \bigl[ 
   \frac{M_Z^2 c_{2 \beta}}{2}
- v_s t_\beta |\lambda| (\frac{|A_{\lambda}|}{\sqrt{2}} c_{\varphi_x} +
|\kappa| \frac{v_s}{2} c_{\varphi_y}) 
+ |\lambda|^2  (\frac{2  s_\beta^2 M_W^2 s_{\theta_W}^2}{e^2} +
\frac{v_s^2}{2})\bigr]~, 
\\
m_{H_u}^2 &= \frac{e}{2 s_\beta M_W s_{\theta_W}}t_{h_u} + \bigl[ \frac{M_Z^2
    c_{2\beta}}{2}  
 + \frac{|\lambda|  v_s}{t_\beta}  (  \frac{|A_{\lambda}|}{\sqrt{2}}
 c_{\varphi_x}+|\kappa| \frac{v_s}{2}  c_{\varphi_y})
- |\lambda|^2 (\frac{2 c_\beta^2 M_W^2 s_{\theta_W}^2}{e^2} +
\frac{v_s^2}{2})\bigr]~,
\\
m_S^2 &= \frac{t_{h_s}}{v_s}  + 
\Bigl[ s_{2 \beta} |\lambda|(\frac{|A_{\lambda}|}{\sqrt{2}}
c_{\varphi_x} +|\kappa| v_s c_{\varphi_y}) 
- |\lambda|^2  v_s\Bigr] \frac{2  M_W^2 s_{\theta_W}^2}{e^2 v_s} -  |\kappa|^2 v_s^2 - 
 \frac{1}{\sqrt{2}} |A_{\kappa}| |\kappa| v_s c_{\varphi_z}~,
\\\nonumber
\varphi_{A_\lambda} &=
\text{sign}_x \Bigl[n_x (\pi) + (-1)^{n_x} 
|\arcsin\Bigl(\frac{e}{\sqrt{2} |A_{\lambda}| M_W s_{\theta_W} s_\beta |\lambda| v_s} t_{a_d} + 
\frac{|\kappa| v_s}{\sqrt{2} |A_\lambda|} s_{\varphi_y}\Bigr)|\Bigr]
 \\& \quad
-\varphi_\lambda - \varphi_s - \varphi_u~, \\\nonumber
\varphi_{A_\kappa} &=  
\text{sign}_z \Bigl[n_z (\pi) + (-1)^{n_z} \nonumber \\& \quad 
|\arcsin \Bigl(\frac{\sqrt{2}}{|A_\kappa| v_s}[\frac{2  M_W s_{\theta_W} c_{\beta}}{e|\kappa| v_s^2 } t_{a_d}
+\frac{3  M_W^2 s_{\theta_W}^2 s_{2 \beta}|\lambda|}{e^2}
  s_{\varphi_y} - \frac{1}{|\kappa| v_s} t_{a_s}] \Bigr)|\Bigr]
-\varphi_\kappa - 3 \varphi_s~, \label{phikappa}\\ 
|A_\lambda| &= \frac{s_{2 \beta}}{\sqrt{2} c_{\varphi_x} |\lambda|v_s c_{\Delta \beta}^2} \Bigl[M_{H^\pm}^2 
- M_W^2 c_{\Delta \beta}^2 -|\kappa||\lambda|v_s^2 \frac{c_{\Delta \beta}^2}{s_{2\beta}} c_{\varphi_y}
+\frac{2 M_W^2 s_{\theta_W}^2 c_{\Delta \beta}^2}{e^2}|\lambda|^2 \nonumber \\&\quad \quad - \frac{e}{2 M_W s_{\theta_W}}
\bigl[t_{h_u} \frac{c_{\beta_c}^2}{s_\beta} + t_{h_d} \frac{s_{\beta_c}^2}{c_\beta}\bigr]\Bigr]~,\\
v_u &= \frac{2 M_W s_{\theta_W} s_\beta}{e}~, \quad 
v_d = \frac{2 M_W s_{\theta_W} c_\beta}{e}~, 
\label{Ersetzung2}
 \\
g &= \frac{e}{s_{\theta_W}}~,\quad g'= \frac{e}{c_{\theta_W}}\quad\ \text{with} \quad\ c_{\theta_W}=\frac{M_W}{M_Z} \quad \text{and} 
\quad s_{\theta_W}^2 = 1 - c_{\theta_W}^2~,
\label{Ersetzung3}
\end{align}
where $n_x$ and $n_z$ can be zero or one in case of two solutions of the tadpole condition, Eqs.~\eqref{tadad} and \eqref{tadas}, and zero if there exists only a single one. Here,
$\text{sign}_x$ and $\text{sign}_z$ are the sign of the corresponding 
arcsine evaluated in the interval $[-\pi, \pi)$, respectively.

\section{Higgs Boson Mass Matrix}
\label{sec:Higgsdetails}

In this section we list the Higgs boson mass matrix elements in a form needed as starting point for the 
renormalization procedure in the basis $\Phi = (h_d, h_u, h_s, A, a_s, G)^T$. This basis is obtained by 
transforming the original basis $\phi =  (h_d, h_u, h_s, a_u, a_d, a_s)^T$ with the matrix
\begin{align}
\mathcal R^{G} = \begin{pmatrix} \id & 0 \\ 0 & \mathcal U^G \end{pmatrix} \quad\ \text{and} \quad
\mathcal U^G = \begin{pmatrix} s_{\beta_n} &  c_{\beta_n} & 0 \\
                               0           &  0           & 1 \\
                               c_{\beta_n} & - s_{\beta_n} & 0 
\end{pmatrix}~. \label{eq:RG}
\end{align}
The angle $\beta_n$ is chosen such that the Goldstone boson field (with zero mass eigenvalue) is 
extracted and, at tree-level, coincides with the angle $\beta$ defined via the ratio of the vacuum 
expectation values, $\beta_n = \beta$. 

The mass matrix elements of the CP-even part $M_{hh}$, \textit{cf.}~Eq.~\eqref{eq:higgsmassmatrix},  are
\begin{align}
\nonumber
M_{h_dh_d} &= \bigl[\frac{M_{H^\pm}^2}{c_{\Delta \beta}^2} - M_W^2 \bigr] s_{\beta}^2 + M_Z^2 c_{\beta}^2 
+ \frac{e c_\beta c_{\beta_B}^2}{2 M_W s_{\theta_W}c_{\Delta \beta}^2}\bigl[(1 + 2 t_{\beta} t_{\beta_B})  t_{h_d} 
                                                  - t_\beta  t_{h_u} \bigr]\\&
\quad\ + 2 |\lambda|^2 M_W^2\frac{s_{\theta_W}^2}{e^2} s_\beta^2~, \\
\nonumber
M_{h_dh_u} &= - \bigl[\frac{M_{H^\pm}^2}{c_{\Delta \beta}^2} - M_W^2 + M_Z^2\bigr] s_{\beta} c_\beta 
+ \frac{e c_\beta c_{\beta_B}^2}{2 M_W s_{\theta_W}c_{\Delta \beta}^2}\bigl[t_{h_u} + t_{\beta} t_{\beta_B}^2 t_{h_d}  \bigr]\\&
\quad\ + |\lambda|^2 M_W^2\frac{s_{\theta_W}^2}{e^2} s_{2 \beta}~,\\
\nonumber
M_{h_uh_u} &= \bigl[\frac{M_{H^\pm}^2}{c_{\Delta \beta}^2} - M_W^2 \bigr] c_{\beta}^2 + M_Z^2 s_{\beta}^2 
+ \frac{e c_\beta s_{2 \beta_B}}{4 M_W s_{\theta_W}c_{\Delta \beta}^2}
      \bigl[(2 + t_\beta t_{\beta_B}) t_{h_u} - t_{\beta_B} t_{h_d}  \bigr]
\\&
\quad\ + 2 |\lambda|^2 M_W^2\frac{s_{\theta_W}^2}{e^2} c_\beta^2~, \\
\nonumber
M_{h_dh_s} &= - \bigl[\frac{M_{H^\pm}^2}{c_{\Delta \beta}^2} - M_W^2 \bigr] \frac{M_W s_{\theta_W} s_\beta s_{2 \beta}}{e v_s}
+ \frac{s_\beta c_\beta c_{\beta_B}^2}{v_s c_{\Delta \beta}^2} \bigl[t_{h_u} + t_{\beta} t_{\beta_B}^2 t_{h_d}  \bigr]
\\&
\quad\  + |\lambda| M_W \frac{s_{\theta_W}}{e}v_s\bigl[2 |\lambda| c_\beta - |\kappa| s_\beta c_{\varphi_y}\bigr]
- \frac{4 |\lambda|^2 M_W^3s_{\theta_W}^3 s_\beta^2 c_\beta}{e^3 v_s}~,\\
\nonumber
M_{h_uh_s} &= - \bigl[\frac{M_{H^\pm}^2}{c_{\Delta \beta}^2} - M_W^2 \bigr] \frac{M_W s_{\theta_W} c_\beta s_{2 \beta}}{e v_s}
+ \frac{c_\beta^2 c_{\beta_B}^2}{v_s c_{\Delta \beta}^2} \bigl[t_{h_u} + t_{\beta} t_{\beta_B}^2 t_{h_d}  \bigr]
\\&
\quad\  + |\lambda| M_W \frac{s_{\theta_W}}{e}v_s\bigl[2 |\lambda| s_\beta - |\kappa| c_\beta c_{\varphi_y}\bigr]
- \frac{4 |\lambda|^2 M_W^3s_{\theta_W}^3 s_\beta c_\beta^2}{e^3 v_s}~,\\
\nonumber
M_{h_sh_s} &= \bigl[\frac{M_{H^\pm}^2}{c_{\Delta \beta}^2} - M_W^2 \bigr] \frac{M_W^2 s_{\theta_W}^2 s_{2 \beta}^2}{e^2 v_s^2}
- \frac{M_W s_{\theta_W} s_{2 \beta} c_\beta c_{\beta_B}^2}{e v_s^2 c_{\Delta \beta}^2} \bigl[t_{h_u} + t_{\beta} t_{\beta_B}^2 t_{h_d}\bigr]
+ \frac{t_{h_s}}{v_s}
\\&
\quad\ + |\lambda| M_W^2 \frac{s_{\theta_W}^2 s_{2 \beta}}{e^2 v_s^2} 
      \bigl[2 |\lambda| M_W^2 \frac{s_{\theta_W}^2}{e^2} s_{2 \beta} -|\kappa|  v_s^2 c_{\varphi_y}\bigr] + 2 |\kappa|^2 v_s^2 
+ \frac{1}{\sqrt{2}} |A_\kappa| |\kappa| v_s c_{\varphi_z}~,
\end{align}
where $\Delta \beta = \beta - \beta_B$ and $\beta_B \equiv \beta_c = \beta_n$. The mixing angle of the 
charged Higgs bosons, $\beta_c$, and the mixing angle $\beta_n$, needed for extracting the Goldstone 
boson, coincide and no discrimination between these two mixing angles is done in the formulae. The angles $\varphi_x$, $\varphi_y$ and $\varphi_z$ have been defined in Eqs. \eqref{eq:varphix}--\eqref{eq:varphiz}. 

The mass matrix elements corresponding to the CP-odd components of the Higgs boson mass matrix are given as
\begin{align}
M_{AA} &= M_{H^\pm}^2 - M_W^2 c_{\Delta \beta}^2 + 2 |\lambda|^2 M_W^2\frac{s_{\theta_W}^2}{e^2} c_{\Delta \beta}^2~, \\
\nonumber
M_{Aa_s} &= \bigl[M_{H^\pm}^2 - M_W^2 c_{\Delta \beta}^2\bigr] \frac{M_W s_{\theta_W} s_{2 \beta}}{e v_s c_{\Delta \beta}} 
- \frac{c_\beta c_{\beta_B}^2}{v_s c_{\Delta \beta}} \bigl[t_{h_u} + t_{\beta} t_{\beta_B}^2 t_{h_d}  \bigr]
\\&
\quad\ + |\lambda| M_W \frac{s_{\theta_W} c_{\Delta \beta}}{e v_s} 
      \bigl[2 |\lambda| M_W^2 \frac{s_{\theta_W}^2}{e^2} s_{2 \beta} - 3 |\kappa| v_s^2 c_{\varphi_y}\bigr]~,
\end{align}
\begin{align}
\nonumber
M_{a_sa_s} &=  \bigl[M_{H^\pm}^2 - M_W^2 c_{\Delta \beta}^2\bigr] \frac{M_W^2 s_{\theta_W}^2 s^2_{2 \beta}}{e^2 v_s^2 c_{\Delta \beta}^2}
- \frac{M_W s_{\theta_W} s_{2 \beta} c_\beta c_{\beta_B}^2}{e v_s^2 c_{\Delta \beta}^2} \bigl[t_{h_u} + t_{\beta} t_{\beta_B}^2 t_{h_d}  \bigr]
+ \frac{t_{h_s}}{v_s}
\\&
\quad\ + |\lambda| M_W^2 \frac{s^2_{\theta_W} s_{2 \beta}}{e^2 v_s^2} 
      \bigl[2 |\lambda| M_W^2 \frac{s_{\theta_W}^2}{e^2} s_{2 \beta} + 3 |\kappa| v_s^2 c_{\varphi_y}\bigr] 
- \frac{3}{\sqrt{2}} |A_\kappa| |\kappa| v_s c_{\varphi_z}~,\\
%
M_{AG} &= \bigl[M_{H^\pm}^2 - M_W^2 c_{\Delta \beta}^2\bigr]t_{\Delta \beta}
+ \frac{e c_{\beta_B}}{2 M_W s_{\theta_W} c_{\Delta \beta}}
      \bigl[ t_{\beta_B} t_{h_d} - t_{h_u}  \bigr] + |\lambda|^2 M_W^2 \frac{s_{\theta_W}^2}{e^2} s_{2 \Delta \beta}~,\\
\nonumber
M_{a_sG} &= \bigl[M_{H^\pm}^2 - M_W^2 c_{\Delta \beta}^2\bigr] \frac{M_W s_{\theta_W} s_{2 \beta}s_{\Delta \beta}}{e v_s c^2_{\Delta \beta}}
- \frac{c_\beta c_{\beta_B}^2 s_{\Delta \beta}}{v_s c^2_{\Delta \beta}} \bigl[t_{h_u} + t_{\beta} t_{\beta_B}^2 t_{h_d}  \bigr]
\\&
\quad\ + |\lambda| M_W \frac{s_{\theta_W} s_{\Delta \beta}}{e v_s} 
      \bigl[2 |\lambda| M_W^2 \frac{s_{\theta_W}^2}{e^2} s_{2 \beta} - 3 |\kappa| v_s^2 c_{\varphi_y}\bigr]~,\\
%
M_{GG} &= \bigl[M_{H^\pm}^2 - M_W^2 c_{\Delta \beta}^2\bigr]t^2_{\Delta \beta} 
+ \frac{e c_{\beta - 2\beta_B}}{2 M_W s_{\theta_W}  c^2_{\Delta \beta}} \bigl[t_{h_d} - t_{\beta - 2\beta_B} t_{h_u}  \bigr]
+ 2 |\lambda|^2 M_W^2\frac{s_{\theta_W}^2}{e^2} s_{\Delta \beta}^2~.
\end{align}
Finally, the mass matrix elements describing the mixing between the CP-even and the CP-odd components 
can be expressed as
\begin{align}
M_{ha} \mathcal {U^G}^T = \begin{pmatrix} \frac{e c_{\beta_B} }{2 M_W s_{\theta_W} s_\beta}t_{a_d} & 
                                          \frac{1}{v_s}t_{a_d} + 3 |\kappa| |\lambda| M_W \frac{s_{\theta_W}}{e} v_s s_\beta s_{\varphi_y} &
                                          - \frac{e s_{\beta_B} }{2 M_W s_{\theta_W} s_\beta} t_{a_d}\\[0.2cm]
                                          \frac{e s_{\beta_B} }{2 M_W s_{\theta_W} s_\beta} t_{a_d}& 
                                          \frac{1}{v_s t_{\beta}} t_{a_d}
                                             + 3 |\kappa| |\lambda| M_W \frac{s_{\theta_W}}{e} v_s  c_\beta s_{\varphi_y} &
                                           \frac{e c_{\beta_B}}{2 M_W s_{\theta_W} s_\beta} t_{a_d}\\[0.2cm]
                                          M_{h_sA} &
                                          M_{h_sa_s} &
                                          M_{h_sG}
                          \end{pmatrix}
\end{align}
with
\begin{align}
 M_{h_sA} &= \frac{ c_{\Delta \beta}}{v_s s_{\beta}}t_{a_d} - 
                                                |\kappa| |\lambda| M_W \frac{s_{\theta_W}}{e} v_s  c_{\Delta \beta} s_{\varphi_y}~,\\
 M_{h_sa_s} &=  \frac{2 }{v_s}t_{a_s} - \frac{4 M_W  s_{\theta_W}}{e}\bigl[\frac{c_\beta}{v_s^2}  t_{a_d}
                                           +  |\kappa| |\lambda|M_W \frac{s_{\theta_W}}{e} s_{2\beta}s_{\varphi_y} \bigr]~,\\
 M_{h_sG} &=   \frac{s_{\Delta \beta} }{v_s s_{\beta}}t_{a_d}
                                           -  |\kappa| |\lambda|M_W \frac{s_{\theta_W}}{e} v_s s_{\Delta \beta}s_{\varphi_y}~.
\end{align}

\section{Chargino and Neutralino Self-Energies}
\label{sec:renselfChaNeu}

In this section the expressions for the renormalized self-energies of the charginos and neutralinos 
are listed. The different parts of the renormalized chargino self-energy decomposed according to Eq.~\eqref{eq:strucself} can be written as ($i,j$=1,2)
\begin{align}
\bigl[\hat{\Sigma}^R_{\chi^+} (p^2)\bigr]_{ij} &= \bigl[\Sigma^R _{\chi^+} (p^2)\bigr]_{ij} +
\frac{1}{2} \bigl[U^* 
(\delta Z_R^C + \delta Z_R^{C^*}) U^T\bigr]_{ij}~, \label{eq:cren1} \\
\bigl[ \hat{\Sigma}^L_{\chi^+} (p^2)\bigr]_{ij} &= \bigl[\Sigma^L _{\chi^+} (p^2)\bigr]_{ij} +
 \frac{1}{2} \bigl[V 
 (\delta Z_L^C + \delta Z_L^{C^*}) V^\dagger\bigr]_{ij}~, \label{eq:cren2} \\
\bigl[\hat{\Sigma}^{Ls}_{\chi^+} (p^2)\bigr]_{ij} &= \bigl[\Sigma^{Ls}_{\chi^+} (p^2)\bigr]_{ij} -
\frac{1}{2}  m_{\chi^\pm_k}
( [U^*\delta Z_R^C  U^T ]_{ik}\,\delta_{kj}
 +  \delta_{ik} [V \delta Z_L^C V^\dagger ]_{kj})  
- \bigl[U^* \delta M_C V^\dagger \bigr]_{ij}~,
\label{eq:cren3} \\
\bigl[\hat{\Sigma}^{Rs}_{\chi^+} (p^2)\bigr]_{ij} &= \bigl[\Sigma^{Rs}_{\chi^+} (p^2) \bigr]_{ij}-
\frac{1}{2}  m_{\chi^\pm_k} 
( [V \delta Z_L^{C^*} V^\dagger]_{ik}\, \delta_{kj} + 
\delta_{ik} [U^* \delta Z_R^{C^*} U^T]_{kj}) - \bigl[V \delta
M_C^\dagger U^T\bigr]_{ij}~,
\label{eq:cren4} 
\end{align}
where, for the renormalization procedure, the chargino spinors as given in Eq.~\eqref{eq:chaspinor}
and the $2\times 2$ chargino mass matrix $M_C$ are replaced by
\begin{align}
\psi^+_{L} &\rightarrow \Big(\id + \frac{1}{2} \delta Z_{L}^C \Big)
\psi^+_{L}~,\\
\psi^-_{R} &\rightarrow \Big(\id + \frac{1}{2} \delta Z_{R}^C \Big)
\psi^-_{R}~,  \quad 
\text{with} \quad
\delta Z_{X}^C = \begin{pmatrix} \delta Z_{X_1}^C & 0 \\
0 & \delta Z_{X_2}^C \end{pmatrix} \quad \text{and}\quad X=L, R\;,
\end{align}
and 
\begin{align}
M_C \to M_C + \delta M_C~,
\end{align}
respectively, where $\delta M_C$ is given in Eq.~\eqref{eq:dMC}.

The various parts of the decomposed renormalized neutralino self-energy can be expressed as ($i,j$=1,...,5)
\begin{align}
\bigl[\hat{\Sigma}^R_{\chi^0} (p^2)\bigr]_{ij} &= \bigl[\Sigma^R_{\chi^0} (p^2)\bigr]_{ij} +
\frac{1}{2} \bigl[{\mathcal N}^* (\delta Z^N
+ \delta Z^{N^*}) {\mathcal N}^T]_{ij}~, \label{eq:nren1} \\
\bigl[ \hat{\Sigma}^L_{\chi^0} (p^2)\bigr]_{ij} &= \bigl[\Sigma^L_{\chi^0} (p^2)\bigr]_{ij} +
 \frac{1}{2} \bigl[{\mathcal N} (\delta Z^N
+ \delta Z^{N^*}) {\mathcal N}^\dagger\bigr]_{ij} \label{eq:nren2}~, \\
\bigl[\hat{\Sigma}^{Ls}_{\chi^0} (p^2) \bigr]_{ij}&= \bigl[\Sigma^{Ls}_{\chi^0} (p^2)\bigr]_{ij} -
\frac{1}{2}m_{\chi^0_k}([{\mathcal N}^* \delta Z^N  {\mathcal N}^\dagger]_{ik} \, \delta_{kj} 
+ \delta_{ik}[{\mathcal N}^* \delta Z^N {\mathcal N}^\dagger]_{kj})  \nonumber \\&\quad
- \bigl[{\mathcal N}^* \delta M_N {\mathcal N}^\dagger  \bigr]_{ij}
\label{eq:nren3}~, \\
\bigl[\hat{\Sigma}^{Rs}_{\chi^0} (p^2)\bigr]_{ij} &= \bigl[\Sigma^{Rs}_{\chi^0} (p^2)\bigr]_{ij} -
\frac{1}{2}  m_{\chi^0_k}([{\mathcal N}\delta Z^{N^*} {\mathcal N}^T ]_{ik}\, \delta_{jk} 
+\delta_{ik} [{\mathcal N}\delta Z^{N^*} {\mathcal N}^T]_{kj}) \nonumber \\&\quad
- \bigl[{\mathcal N} \delta M_N^\dagger {\mathcal N}^T \bigr]_{ij}~.
\label{eq:nren4} 
\end{align}
For the renormalization procedure, the neutralino spinor defined in Eq.~\eqref{eq:neuspinor} has been replaced by
\begin{align}
\psi^0 &\rightarrow \Big(\id + \frac{1}{2} \delta Z^N \Big)
\psi^0 \quad \text{with} \quad 
\delta Z^N = \text{diag}(\delta Z^N_1,\,\delta Z^N_2,\,\delta Z^N_3,\,\delta Z^N_4,\, \delta Z^N_5)
\end{align}
and the neutralino mass matrix by
\begin{align}
M_N \rightarrow M_N + \delta M_N~.
\end{align}
The matrix elements of $\delta M_N$ can be found in Eqs.~\eqref{eq:dMN11}--\eqref{eq:dMNzero}.
\end{appendix}


\begin{thebibliography}{10}
%
\bibitem{cmshiggs}
S.~Chatrchyan {\it et al.}  [CMS Collaboration],
  arXiv:1202.1416 [hep-ex],
  arXiv:1202.1487 [hep-ex],
  arXiv:1202.1488 [hep-ex], 
  arXiv:1202.1489 [hep-ex],
  arXiv:1202.1997 [hep-ex],
  arXiv:1202.3478 [hep-ex],
  arXiv:1202.3617 [hep-ex],
  arXiv:1202.4083 [hep-ex] and
  arXiv:1202.4195 [hep-ex].
%
\bibitem{atlashiggs}
G.~Aad {\it et al.}  [ATLAS Collaboration],
  Phys.\ Lett.\ B {\bf 710} (2012) 49
  [arXiv:1202.1408 [hep-ex]], 
Phys.\ Rev.\ Lett.\  {\bf 108}, 111803 (2012)
  [arXiv:1202.1414 [hep-ex]],
  Phys.\  Lett.\ B {\bf 710} (2012) 383
  [arXiv:1202.1415 [hep-ex]],
  arXiv:1206.0756 [hep-ex],
  arXiv:1206.2443 [hep-ex],
  arXiv:1206.5971 [hep-ex],
  arXiv:1206.6074 [hep-ex] and
  arXiv:1205.6744 [hep-ex].

%
\bibitem{susy}
D.V.~Volkov and V.P.~Alkulov, Phys. Lett. {\bf B46} (1973) 109;
J.~Wess and B.~Zumino, Nucl. Phys. {\bf B70} (1974) 39; 
P.~Fayet, Phys. Lett. {\bf B64} (1976) 159, Phys. Lett. {\bf B69}
(1977) 489, Phys. Lett. {\bf B84} (1979) 416; G.F. Farrar and
P. Fayet, Phys. Lett. {\bf B76} (1978) 575; S. Dimopoulos and
H. Georgi, Nucl. Phys. {\bf B193}  (1981) 150; N. Sakai, Z. Phys. {\bf
C11} (1981) 153; E.~Witten,
Nucl. Phys. {\bf B188} (1981) 513; H.P.~Nilles, Phys. Rep. {\bf 110} (1984) 1; 
H.E.~Haber and G.L.~Kane, Phys. Rep. {\bf 117} (1985) 75;
M.F.~Sohnius, Phys. Rep. {\bf 128} (1985) 39; J.F. Gunion
and H.E. Haber, Nucl. Phys. {\bf B272} (1986) 1 [Erratum-ibid. {\bf
  B402} (1993) 567], Nucl. Phys. {\bf B278} (1986) 449;  A.B. Lahanas
and D.V. Nanopoulos, Phys. Rep. {\bf 145} (1987) 1.
%
\bibitem{mssm}
For reviews and further references, see: J.F. Gunion, H.E. Haber, G. Kane
  and S. Dawson, {\it ``The Higgs Hunter's Guide''}, Addison-Wesley,
  1990; S.P. Martin, [hep-ph/9709356];
S.~Dawson,
  [hep-ph/9712464];
 M.~Gomez-Bock, M.~Mondragon, M.~M\"uhlleitner, R.~Noriega-Papaqui,
  I.~Pedraza, M.~Spira and P.~M.~Zerwas, 
  J.\ Phys.\ Conf.\ Ser.\  {\bf 18} (2005) 74
  [arXiv:hep-ph/0509077];
  M.~Gomez-Bock, M.~Mondragon, M.~M\"uhlleitner, M.~Spira and P.~M.~Zerwas,
  [arXiv:0712.2419 [hep-ph]];
  A.~Djouadi,
  Phys.\ Rept.\  {\bf 459} (2008)  1
  [hep-ph/0503173].
%
\bibitem{nmssm}
P.~Fayet,
  Nucl.\ Phys.\  {\bf B90} (1975)  104;
R.~Barbieri, S.~Ferrara, C.~A.~Savoy,
  Phys.\ Lett.\  {\bf B119} (1982)  343;
M.~Dine, W.~Fischler, M.~Srednicki,
  Phys.\ Lett.\  {\bf B104} (1981)  199;
H.~P.~Nilles, M.~Srednicki, D.~Wyler,
  Phys.\ Lett.\  {\bf B120} (1983)  346;
J.~M.~Frere, D.~R.~T.~Jones, S.~Raby,
  Nucl.\ Phys.\  {\bf B222} (1983)  11;
J.~P.~Derendinger, C.~A.~Savoy,
  Nucl.\ Phys.\  {\bf B237} (1984)  307;
%
J.~R.~Ellis, J.~F.~Gunion, H.~E.~Haber, L.~Roszkowski, F.~Zwirner,
  Phys.\ Rev.\  {\bf D39} (1989)  844;
M.~Drees,
  Int.\ J.\ Mod.\ Phys.\  {\bf A4} (1989)  3635;
U.~Ellwanger, M.~Rausch de Traubenberg, C.~A.~Savoy,
  Phys.\ Lett.\  {\bf B315} (1993)  331 [hep-ph/9307322],
  Z.\ Phys.\  {\bf C67} (1995)  665 [hep-ph/9502206],
  Nucl.\ Phys.\  {\bf B492} (1997)  21 [hep-ph/9611251];
T.~Elliott, S.~F.~King, P.~L.~White,
  Phys.\ Lett.\  {\bf B351} (1995)  213 [hep-ph/9406303];
S.~F.~King, P.~L.~White,
  Phys.\ Rev.\  {\bf D52} (1995)  4183 [hep-ph/9505326];
F.~Franke, H.~Fraas,
  Int.\ J.\ Mod.\ Phys.\  {\bf A12} (1997)  479 [hep-ph/9512366];
%
M.~Maniatis,
  Int.\ J.\ Mod.\ Phys.\  {\bf A25} (2010) 3505 [arXiv:0906.0777
  [hep-ph]];
U.~Ellwanger, C.~Hugonie, A.~M.~Teixeira,
  Phys.\ Rept.\  {\bf 496} (2010) 1 [arXiv:0910.1785 [hep-ph]].
%
\bibitem{muproblem}
J.E. Kim and H.P. Nilles, Phys. Lett. {\bf B138} (1984) 150.
%
\bibitem{finetune}
  M.~Bastero-Gil, C.~Hugonie, S.~F.~King, D.~P.~Roy and S.~Vempati,
  Phys.\ Lett.\ B\ {\bf 489}, 359  (2000)
  [hep-ph/0006198];
  A.~Delgado, C.~Kolda, J.~P.~Olson and A.~de la Puente,
  Phys.\ Rev.\ Lett.\ \ {\bf 105}, 091802  (2010)
  [arXiv:1005.1282 [hep-ph]];
U.~Ellwanger, G.~Espitalier-Noel and C.~Hugonie,
  JHEP {\bf 1109} (2011) 105
  [arXiv:1107.2472 [hep-ph]];
  G.~G.~Ross and K.~Schmidt-Hoberg,
  arXiv:1108.1284 [hep-ph].

\bibitem{King:2012is}
  S.~F.~King, M.~Muhlleitner and R.~Nevzorov,
  Nucl.\ Phys.\ B {\bf 860} (2012) 207
  [arXiv:1201.2671 [hep-ph]].

%
\bibitem{brchanges0}
  L.~J.~Hall, D.~Pinner and J.~T.~Ruderman,
  JHEP {\bf 1204} (2012) 131
  [arXiv:1112.2703 [hep-ph]].

\bibitem{brchanges}
U.~Ellwanger,
  Phys.\ Lett.\ B {\bf 698} (2011) 293
  [arXiv:1012.1201 [hep-ph]];
U.~Ellwanger,
  JHEP {\bf 1203} (2012) 044
  [arXiv:1112.3548 [hep-ph]];
A.~Arvanitaki and G.~Villadoro,
  JHEP {\bf 1202} (2012) 144
  [arXiv:1112.4835 [hep-ph]].
%
\bibitem{radcpmssm0}
E.~Accomando {\it et al.},
  hep-ph/0608079.
%
\bibitem{radcpmssm}
A.~Pilaftsis,
  Phys.\ Lett.\ B {\bf 435} (1998) 88
  [hep-ph/9805373];
D.~A.~Demir,
  Phys.\ Rev.\ D {\bf 60} (1999) 055006 [hep-ph/9901389];
A.~Pilaftsis and C.~E.~M.~Wagner,
  Nucl.\ Phys.\ B {\bf 553} (1999) 3 [hep-ph/9902371];
S.~Y.~Choi, M.~Drees and J.~S.~Lee,
  Phys.\ Lett.\ B {\bf 481} (2000) 57 [hep-ph/0002287];
M.~S.~Carena, J.~R.~Ellis, A.~Pilaftsis and C.~E.~M.~Wagner,
  Nucl.\ Phys.\ B {\bf 586} (2000) 92 [hep-ph/0003180];
T.~Ibrahim and P.~Nath,
  Phys.\ Rev.\ D {\bf 63} (2001) 035009 [hep-ph/0008237] and
  Phys.\ Rev.\ D {\bf 66} (2002) 015005 [hep-ph/0204092];
S.~Heinemeyer,
  Eur.\ Phys.\ J.\ C {\bf 22} (2001) 521 [hep-ph/0108059];
M.~S.~Carena, J.~R.~Ellis, A.~Pilaftsis and C.~E.~M.~Wagner,
  Nucl.\ Phys.\ B {\bf 625} (2002) 345 [hep-ph/0111245];
S.~W.~Ham, S.~K.~Oh, E.~J.~Yoo, C.~M.~Kim and D.~Son,
  Phys.\ Rev.\ D {\bf 68} (2003) 055003 [hep-ph/0205244];
M.~Frank, S.~Heinemeyer, W.~Hollik and G.~Weiglein,
  hep-ph/0212037;
S.~Heinemeyer,
  Int.\ J.\ Mod.\ Phys.\ A {\bf 21} (2006) 2659 [hep-ph/0407244];
S.~Heinemeyer, W.~Hollik, H.~Rzehak and G.~Weiglein,
  Phys.\ Lett.\ B {\bf 652} (2007) 300 [arXiv:0705.0746 [hep-ph]].
%
\bibitem{1lfull}
M.~Frank, T.~Hahn, S.~Heinemeyer, W.~Hollik, H.~Rzehak, G.~Weiglein,
  JHEP {\bf 0702 } (2007)  047 [hep-ph/0611326].
%
\bibitem{mssmcpveff}
E.~Christova, H.~Eberl, W.~Majerotto and S.~Kraml,
 Nucl.\ Phys.\ B {\bf 639} (2002) 263 [Erratum-ibid.\ B {\bf 647}
 (2002) 359] [hep-ph/0205227]; 
E.~Christova, H.~Eberl, W.~Majerotto and S.~Kraml,
 JHEP {\bf 0212} (2002) 021 [hep-ph/0211063];
.~Khater and P.~Osland,
 Nucl.\ Phys.\ B {\bf 661} (2003) 209 [hep-ph/0302004];
 S.~Y.~Choi, J.~Kalinowski, Y.~Liao and P.~M.~Zerwas,
  Eur.\ Phys.\ J.\ C {\bf 40} (2005) 555 [hep-ph/0407347];
K.~E.~Williams and G.~Weiglein,
  Phys.\ Lett.\ B {\bf 660} (2008) 217 [arXiv:0710.5320 [hep-ph]];
K.~E.~Williams, H.~Rzehak and G.~Weiglein,
  Eur.\ Phys.\ J.\ C {\bf 71} (2011) 1669 [arXiv:1103.1335 [hep-ph]].
%
%
\bibitem{escapedet}
M.~S.~Carena, J.~R.~Ellis, A.~Pilaftsis and C.~E.~M.~Wagner,
  Phys.\ Lett.\ B {\bf 495} (2000) 155
  [hep-ph/0009212];
M.~S.~Carena, J.~R.~Ellis, S.~Mrenna, A.~Pilaftsis and C.~E.~M.~Wagner,
  Nucl.\ Phys.\ B {\bf 659} (2003) 145 [hep-ph/0211467];
A.~Dedes and S.~Moretti,
  Phys.\ Rev.\ Lett.\  {\bf 84} (2000) 22 [hep-ph/9908516];
A.~Dedes and S.~Moretti,
  Nucl.\ Phys.\ B {\bf 576} (2000) 29
  [hep-ph/9909418];
S.~Y.~Choi and J.~S.~Lee,
  Phys.\ Rev.\ D {\bf 61} (2000) 115002 [hep-ph/9910557];
S.~Y.~Choi, K.~Hagiwara and J.~S.~Lee,
  Phys.\ Lett.\ B {\bf 529} (2002) 212 [hep-ph/0110138];
A.~Arhrib, D.~K.~Ghosh and O.~C.~W.~Kong,
  Phys.\ Lett.\ B {\bf 537} (2002) 217 [hep-ph/0112039];
B.~E.~Cox, J.~R.~Forshaw, J.~S.~Lee, J.~Monk and A.~Pilaftsis,
  Phys.\ Rev.\ D {\bf 68} (2003) 075004 [hep-ph/0303206];
A.~G.~Akeroyd,
  Phys.\ Rev.\ D {\bf 68} (2003) 077701 [hep-ph/0306045];
F.~Borzumati, J.~S.~Lee and W.~Y.~Song,
  Phys.\ Lett.\ B {\bf 595} (2004) 347 [hep-ph/0401024];
V.~A.~Khoze, A.~D.~Martin and M.~G.~Ryskin,
  Eur.\ Phys.\ J.\ C {\bf 34} (2004) 327 [hep-ph/0401078;
J.~R.~Ellis, J.~S.~Lee and A.~Pilaftsis,
  Phys.\ Rev.\ D {\bf 70} (2004) 075010 [hep-ph/0404167];
D.~K.~Ghosh, R.~M.~Godbole and D.~P.~Roy,
  Phys.\ Lett.\ B {\bf 628} (2005) 131 [hep-ph/0412193];
D.~K.~Ghosh and S.~Moretti,
  Eur.\ Phys.\ J.\ C {\bf 42} (2005) 341 [hep-ph/0412365];
J.~R.~Ellis, J.~S.~Lee and A.~Pilaftsis,
  Phys.\ Rev.\ D {\bf 71} (2005) 075007 [hep-ph/0502251].
%
%
\bibitem{spontcpviol}
J.~C.~Romao,
  Phys.\ Lett.\ B {\bf 173} (1986) 309.
%
\bibitem{cpviolconstr}
B.~C.~Regan, E.~D.~Commins, C.~J.~Schmidt and D.~DeMille,
  Phys.\ Rev.\ Lett.\  {\bf 88} (2002) 071805;
C.~A.~Baker {\it et al.},
  Phys.\ Rev.\ Lett.\  {\bf 97} (2006) 131801 [hep-ex/0602020];
W.~C.~Griffith {\it et al.},
  Phys.\ Rev.\ Lett.\  {\bf 102} (2009) 101601.
%
\bibitem{cpphaseok}
M.~Matsuda and M.~Tanimoto,
  Phys.\ Rev.\ D {\bf 52} (1995) 3100 [hep-ph/9504260];
N.~Haba,
  Prog.\ Theor.\ Phys.\  {\bf 97} (1997) 301 [hep-ph/9608357];
T.~Ibrahim and P.~Nath,
  Phys.\ Rev.\ D {\bf 58} (1998) 111301
   [Erratum-ibid.\ D {\bf 60} (1999) 099902] [hep-ph/9807501];
J.~R.~Ellis, J.~S.~Lee and A.~Pilaftsis,
  JHEP {\bf 0810} (2008) 049 [arXiv:0808.1819 [hep-ph]];
M.~Boz,
  Mod.\ Phys.\ Lett.\ A {\bf 21} (2006) 243
  [hep-ph/0511072].
%
\bibitem{ewbg}
K.~Cheung, T.~-J.~Hou, J.~S.~Lee and E.~Senaha,
  Phys.\ Rev.\ D {\bf 84} (2011) 015002 [arXiv:1102.5679 [hep-ph]].
%
\bibitem{garisto}
R.~Garisto,
  Phys.\ Rev.\ D {\bf 49} (1994) 4820 [hep-ph/9311249].
%
\bibitem{squarkrad}
S.~W.~Ham, J.~Kim, S.~K.~Oh and D.~Son,
  Phys.\ Rev.\ D {\bf 64} (2001) 035007 [hep-ph/0104144];
S.~W.~Ham, S.~H.~Kim, S.~K.~OH and D.~Son,
  Phys.\ Rev.\ D {\bf 76} (2007) 115013 [arXiv:0708.2755 [hep-ph]].
%
\bibitem{radall}
S.~W.~Ham, S.~K.~Oh and D.~Son,
  Phys.\ Rev.\ D {\bf 65} (2002) 075004 [hep-ph/0110052];
S.~W.~Ham, Y.~S.~Jeong and S.~K.~Oh,
  hep-ph/0308264.

\bibitem{Funakubo:2004ka}
  K.~Funakubo and S.~Tao,
  Prog.\ Theor.\ Phys.\  {\bf 113} (2005) 821
  [hep-ph/0409294].

\bibitem{cpeff2loop}
K.~Cheung, T.~-J.~Hou, J.~S.~Lee and E.~Senaha,
  Phys.\ Rev.\ D {\bf 82} (2010) 075007 [arXiv:1006.1458 [hep-ph]].
%
\bibitem{horealnmssm}
U.~Ellwanger,
  Phys.\ Lett.\  {\bf B303} (1993)  271 [hep-ph/9302224];
T.~Elliott, S.~F.~King, P.~L.~White,
  Phys.\ Lett.\  {\bf B305} (1993) 71 [hep-ph/9302202],
  Phys.\ Lett.\  {\bf B314} (1993)  56 [hep-ph/9305282],
  Phys.\ Rev.\  {\bf D49} (1994)  2435 [hep-ph/9308309];
P.~N.~Pandita,
  Z.\ Phys.\  {\bf C59} (1993)  575,
  Phys.\ Lett.\  {\bf B318} (1993)  338;
U.~Ellwanger, C.~Hugonie,
  Phys.\ Lett.\  {\bf B623} (2005)  93 [hep-ph/0504269];
G.~Degrassi, P.~Slavich,
  Nucl.\ Phys.\  {\bf B825} (2010)  119 [arXiv:0907.4682 [hep-ph]];
F.~Staub, W.~Porod, B.~Herrmann,
  JHEP {\bf 1010} (2010)  040 [arXiv:1007.4049 [hep-ph]].
\bibitem{realnmssm}
K.~Ender, T.~Graf, M.~Muhlleitner and H.~Rzehak,
  Phys.\ Rev.\ D {\bf 85} (2012) 075024
  [arXiv:1111.4952 [hep-ph]].

\bibitem{LEPHbb}
R.~Barate {\it et al.}  [LEP Working Group for Higgs boson searches],
Phys.\ Lett.\  B {\bf 565} (2003) 61
[hep-ex/0306033];
S.~Schael {\it et al.}  [ALEPH and DELPHI and L3 and OPAL Collaborations],
Eur.\ Phys.\ J.\  C {\bf 47} (2006) 547
[hep-ex/0602042].

\bibitem{tevsearch}
[TEVNPH (Tevatron New Phenomena and Higgs Working Group) and CDF and D0 Collaborations],
  arXiv:1203.3774 [hep-ex].
%
\bibitem{CKM}
M.~Kobayashi and T.~Maskawa,
  Prog.\ Theor.\ Phys.\  {\bf 49} (1973) 652.

\bibitem{Denner:1991kt}
  A.~Denner,
  Fortsch.\ Phys.\  {\bf 41} (1993) 307
  [arXiv:0709.1075 [hep-ph]].

\bibitem{vuvd}
  A.~Brignole,
  Phys.\ Lett.\ B {\bf 281} (1992) 284,
P.~H.~Chankowski, S.~Pokorski, J.~Rosiek;
  Phys.\ Lett.\  {\bf B286} (1992)  307, 
  Nucl.\ Phys.\  {\bf B423} (1994)  437 [hep-ph/9303309];
A.~Dabelstein,
  Z.\ Phys.\  {\bf C67} (1995)  495 [hep-ph/9409375],
  Nucl.\ Phys.\  {\bf B456} (1995)  25 [hep-ph/9503443];
A.~Freitas and D.~Stockinger,
  Phys.\ Rev.\ D {\bf 66} (2002) 095014
  [hep-ph/0205281].

\bibitem{Frank:2003tg}
  M.~Frank,
  Berlin, Germany: RHOMBOS-Verl. (2003) 148 p.
\bibitem{Hollik:2002mv}
  W.~Hollik, E.~Kraus, M.~Roth, C.~Rupp, K.~Sibold and D.~Stockinger,
  Nucl.\ Phys.\ B {\bf 639} (2002) 3
  [hep-ph/0204350].

\bibitem{feynarts}
J.~Kublbeck, M.~Bohm, A.~Denner,
  Comput.\ Phys.\ Commun.\  {\bf 60} (1990)  165;
  T.~Hahn,
  Comput.\ Phys.\ Commun.\  {\bf 140} (2001)  418 [hep-ph/0012260];
T.~Hahn, C.~Schappacher,
  Comput.\ Phys.\ Commun.\  {\bf 143} (2002)  54 [hep-ph/0105349].

\bibitem{sarah}
F.~Staub,
 [arXiv:0806.0538 [hep-ph]],
%
Comput.\ Phys.\ Commun.\  {\bf 181} (2010)  1077 [arXiv:0909.2863
[hep-ph]], 
%
Comput.\ Phys.\ Commun.\  {\bf 182} (2011)  808 [arXiv:1002.0840
[hep-ph]].

\bibitem{formcalc}
 T.~Hahn, M.~Perez-Victoria,
  Comput.\ Phys.\ Commun.\  {\bf 118} (1999)  153 [hep-ph/9807565];
T.~Hahn,
  Comput.\ Phys.\ Commun.\  {\bf 178} (2008)  217 [hep-ph/0611273].

\bibitem{constrained}
 F.~del Aguila, A.~Culatti, R.~Munoz Tapia, M.~Perez-Victoria,
 Nucl.\ Phys.\  {\bf B537} (1999)  561 [hep-ph/9806451].

\bibitem{slha}
P.~Z.~Skands, B.~C.~Allanach, H.~Baer, C.~Balazs, G.~Belanger, F.~Boudjema, A.~Djouadi, R.~Godbole {\it et al.},
  JHEP {\bf 0407} (2004) 036 [hep-ph/0311123];
B.~C.~Allanach, C.~Balazs, G.~Belanger, M.~Bernhardt, F.~Boudjema, D.~Choudhury, K.~Desch, U.~Ellwanger {\it et al.},
  Comput.\ Phys.\ Commun.\  {\bf 180} (2009)  8 [arXiv:0801.0045
  [hep-ph]];
 F.~Mahmoudi, S.~Heinemeyer, A.~Arbey, A.~Bharucha, T.~Goto, T.~Hahn, U.~Haisch, S.~Kraml {\it et al.},
  [arXiv:1008.0762 [hep-ph]].

\bibitem{nmssmtools}
U.~Ellwanger, J.F.~Gunion and C.~Hugonie, JHEP {\bf 0502}
(2005) 066;
U.~Ellwanger and C.~Hugonie, 
Comput.\ Phys.\ Commun.\  {\bf 175} (2006) 290;
U.~Ellwanger and C.~Hugonie, Comput.\ Phys.\ Commun.\  {\bf 177}
(2007) 399; \\
(see also {\tt
  http://www.th.u-psud.fr/NMHDECAY/nmssmtools.html}). 

\bibitem{pdg}
K. Nakamura et al. (Particle Data Group), J. Phys. {\bf G37} (2010) 075021
and 2011 partial update for the 2012 edition. 


\bibitem{higgsbounds}
P.~Bechtle, O.~Brein, S.~Heinemeyer, G.~Weiglein and K.~E.~Williams,
  Comput.\ Phys.\ Commun.\  {\bf 181} (2010) 138
  [arXiv:0811.4169 [hep-ph]];
P.~Bechtle, O.~Brein, S.~Heinemeyer, G.~Weiglein and K.~E.~Williams,
  Comput.\ Phys.\ Commun.\  {\bf 182} (2011) 2605
  [arXiv:1102.1898 [hep-ph]].

\bibitem{higlu}
 M.~Spira,
  ``HIGLU: A Program for the Calculation of the Total Higgs Production 
Cross-Section at Hadron Colliders via Gluon Fusion including QCD 
Corrections,''
  [hep-ph/9510347].

\bibitem{programs}
URL: {\tt http://people.web.psi.ch/spira/proglist.html}

\bibitem{nloqcdtth}
 W.~Beenakker, S.~Dittmaier, M.~Kramer, B.~Plumper, M.~Spira and 
P.~M.~Zerwas,
  Phys.\ Rev.\ Lett.\  {\bf 87} (2001) 201805
  [hep-ph/0107081];
  Nucl.\ Phys.\  B {\bf 653} (2003) 151
  [hep-ph/0211352];
L. Reina and S. Dawson, 
Phys. Rev. Lett. {\bf 87} (2001) 201804, arXiv:hep-ph/0107101;
S.~Dawson, L.~H.~Orr, L.~Reina and D.~Wackeroth,
  Phys.\ Rev.\  D {\bf 67}, 071503 (2003)
  [hep-ph/0211438].

\bibitem{higgswg}
S.~Dittmaier {\it et al.}  [LHC Higgs Cross Section Working Group
Collaboration], 
  arXiv:1101.0593 [hep-ph]; \newline
URL: {\tt
  https://twiki.cern.ch/twiki/bin/view/LHCPhysics/CrossSections}

\bibitem{hdecay}
A. Djouadi, M. Spira and P.M. Zerwas, Phys. Lett.  B {\bf 264}
(1991) 440 and  Z. Phys. C {\bf 70} (1996) 427;  M. Spira {\it et al.}, Nucl.\
Phys.\ B {\bf 453} (1995) 17;  A.~Djouadi, J.~Kalinowski and M.~Spira, Comput.
Phys. Commun. {\bf 108} (1998) 56;
A. Djouadi, J. Kalinowski, M. Muhlleitner and M. Spira in 
J.~M.~Butterwort {\it et al.},
  arXiv:1003.1643 [hep-ph].
%
\bibitem{susyhit}
A.~Djouadi, M.~M.~Muhlleitner and M.~Spira,
  Acta Phys.\ Polon.\ {\bf B38} (2007) 635
  [hep-ph/0609292].

\bibitem{atlassquark}
G.~Aad {\it et al.}  [ATLAS Collaboration],
  Phys.\ Lett.\ B {\bf 701} (2011) 186
  [arXiv:1102.5290 [hep-ex]];
  Phys.\ Lett.\ B {\bf 701} (2011) 398
  [arXiv:1103.4344 [hep-ex]];
G.~Aad {\it et al.}  [ATLAS Collaboration],
  Phys.\ Lett.\ B {\bf 710} (2012) 67
  [arXiv:1109.6572 [hep-ex]].
G.~Aad {\it et al.}  [ATLAS Collaboration],
  Phys.\ Rev.\ D {\bf 85} (2012) 012006
  [arXiv:1109.6606 [hep-ex]];
G.~Aad {\it et al.}  [Atlas Collaboration],
  JHEP {\bf 1111} (2011) 099
  [arXiv:1110.2299 [hep-ex]];
G.~Aad {\it et al.}  [ATLAS Collaboration],
  Phys.\ Rev.\ Lett.\  {\bf 108} (2012) 181802
  [arXiv:1112.3832 [hep-ex]].

\bibitem{cmssquark}
 S.~Chatrchyan {\it et al.}  [CMS Collaboration],
 Phys. Rev. Lett. {\bf 107} (2011) 221804
  [arXiv:1109.2352 [hep-ex]];
The CMS Collaboration, CMS-PAS-SUS-11-002, CMS-PAS-SUS-11-005,
CMS-PAS-SUS-11-010, CMS-PAS-SUS-11-016.
 
\bibitem{micromegas}
G.~Belanger, F.~Boudjema, A.~Pukhov and A.~Semenov,
Comput.\ Phys.\ Commun.\  {\bf 149} (2002) 103 [hep-ph/0112278],  
Comput.\ Phys.\ Commun.\ {\bf 174} (2006) 577 [hep-ph/0405253] and
Comput.\ Phys.\ Commun.\  {\bf 180} (2009) 747 [arXiv:0803.2360 [hep-ph]];
G.~Belanger et al.,
  Comput.\ Phys.\ Commun.\  {\bf 182} (2011) 842
  [arXiv:1004.1092 [hep-ph]].

\bibitem{thirdgenatlas}
G.~Aad {\it et al.}  [ATLAS Collaboration],
  Phys.\ Lett.\ B {\bf 701} (2011) 1
  [arXiv:1103.1984 [hep-ex]];
G.~Aad {\it et al.}  [ATLAS Collaboration],
  [arXiv:1203.6193 [hep-ex]];
G.~Aad {\it et al.}  [ATLAS Collaboration],
  arXiv:1204.6736 [hep-ex].

\bibitem{thirdgencms}
S.~Chatrchyan {\it et al.}  [CMS Collaboration],
  arXiv:1205.3933 [hep-ex];
The CMS Collaboration, CMS-PAS-SUS-11-027, CMS-PAS-SUS-11-028. 

\end{thebibliography}
\end{document}